\documentstyle[preprint,aps]{revtex}
\begin{document}
\tighten
\draft
\title{Critical analysis of quark-meson coupling models for nuclear matter
       and finite nuclei}
\author{H.~M\"uller and B.~K.~Jennings}
\address{TRIUMF, 4004 Wesbrook Mall, Vancouver, B.C. Canada V6T 2A3}
\vskip1in
\date{\today}
\maketitle
\begin{abstract}
Three versions of the quark-meson coupling (QMC) model are applied to 
describe properties of nuclear matter and finite nuclei.
The models differ in the treatment of the bag constant and in terms
of nonlinear scalar self-interactions.
In two versions of the model the bag constant is held fixed at its free space
value whereas in the third model it depends on the density of the
nuclear environment. As a consequence opposite predictions for 
the medium modifications of the internal nucleon structure arise.
After calibrating the model parameters at equilibrium nuclear matter density,
binding energies, charge radii, single-particle spectra and density 
distributions of spherical nuclei are analyzed and compared with QHD
calculations. For the models which predict a decreasing size of the nucleon 
in the nuclear environment, unrealistic features of the nuclear shapes arise.
\end{abstract}
\vspace{20pt}
\pacs{PACS number(s): 12.39.Ba, 21.60.-n, 21.10.-k, 24.85.+p}
%%%%%%%%%%%%%%%%%%%%%%%%%%%%%%%%%%%%%%%%%%%%%%%%%%%%%%%%%%%%%%%%%%%
%%%%%%%%%%%%%%%%%%%%%%%%%%%%%%%%%%%%%%%%%%%%%%%%%%%%%%%%%%%%%%%%%%%
%
\section{Introduction}
Although quantum chromodynamics is believed to be the fundamental
theory of strong interactions, low- and medium-energy nuclear phenomenology 
is successfully described in terms of hadronic degrees of freedom.
Theoretical challenges arise in phenomena which reveal the quark structure 
of hadrons.
Building models which connect observed nuclear phenomena and the underlying
physics of strong interactions has become one of the fundamental
goals in nuclear physics.
Such models are necessarily crude and contain a variety of unknown 
parameters since the study of the nuclear many-body problem on the fundamental 
level is intractable. However, it is important that the new models 
respect established results which are successfully described in
the hadronic framework.

The quark-meson coupling (QMC) model proposed by Guichon \cite{GUICHON88}, 
provides a simple framework
to incorporate quark degrees of freedom in the study of nuclear many-body
systems.
In the QMC model nucleons arise as non-overlapping 
MIT bags interacting through meson mean fields.
The model has been applied to a variety of problems in nuclear physics.
It was shown that it describes the saturation properties of nuclear
matter \cite{FLECK90,SAITO94a,SAITO95b,JIN96a,JIN96b,MUELLER97},
and that it gives a fair description of the bulk properties of
finite nuclei \cite{GUICHON96,SAITO96,BLUNDEN96,SAITO97,GUICHON97,MUELLER98}.
The model was extended to include hyperons \cite{SAITO95a} and applied
to studies of hyper-nuclei \cite{TSUSHIMA97a,TSUSHIMA97b}.

Although it provides a simple and intuitive framework to describe 
the basic features of nuclear systems in terms of quark degrees of freedom, 
the QMC model has a serious shortcoming. 
It predicts much smaller scalar and vector potentials than 
obtained in successful hadronic models \cite{SAITO94a,JIN96a,JIN96b}.
As a consequence the nucleon mass is too high and the 
spin-orbit force is too weak to explain spin-orbit splittings 
in finite nuclei \cite{SAITO96,BLUNDEN96,SAITO97}.

The QMC model can be significantly improved by introducing the concept of
a density dependent bag constant \cite{JIN96a,JIN96b}.
The modified quark-meson coupling (MQMC) produces large scalar and vector 
potentials under the condition that the value of the bag constant in 
the nuclear environment significantly drops below its free-space value. 

In the present work we study three different versions of the QMC model.
The first model is the original QMC model which we denote by QMCI.
The second model denoted by QMCII is an extension which includes cubic and 
quartic self interactions for the scalar meson.
As our third model we adopt model ${\rm MQMC}_{\rm A}$ from 
Ref.~\cite{MUELLER97}, in which the bag constant is a function of the scalar 
field.

The original idea of the QMC model was to calibrate the
model in free space such that the nucleon mass is reproduced
and then to extrapolate to many-nucleon systems. 
However, it is not possible
to account for all the necessary degrees of freedom. Most importantly, the
scalar field which describes the mid-range part of the nucleon-nucleon
interaction cannot be described in the framework 
of the simple bag model, but it must be included. 
The additional cubic and quartic scalar self-interactions of the QMCII model
were introduced in Refs.~\cite{SAITO97} to model the
density dependence of the scalar mass.
The corresponding couplings cannot be determined on a 
fundamental level \cite{SAITO97}.
In our approach we choose these parameters to reproduce the
properties of nuclear matter near equilibrium that are 
known to be characteristic of the observed bulk and single-particle properties 
of nuclei. This procedure guarantees the reproduction of large scalar and
vector potentials. 

In the MQMC model new parameters arise from the unknown density dependence 
of the bag constant.
A refined MQMC model can be accurately calibrated to produce 
the empirical saturation properties of nuclear matter \cite{MUELLER97}
and it provides a good description of the bulk properties of finite 
nuclei \cite{MUELLER98}.
Most importantly, the model can be calibrated to reproduce small values for 
the effective nucleon mass which leads to a realistic description of spin-orbit 
splittings \cite{MUELLER98}.

One of the basic motivations of applying quark models to nuclear systems is the
hope to describe medium modifications of the internal quark-substructure
of the nucleon.
An important phenomenological quantity is the nucleon size which is 
represented in the framework of the QMC model by the bag radius \cite{JIN97}.
Due to the increase of the bag constant the MQMC model predicts significantly 
swollen nucleons in the nuclear environment \cite{JIN96a,JIN96b,MUELLER97}. 
In contrast, for the QMCI and II model the bag radius decreases with 
increasing density\cite{SAITO94a,GUICHON96,SAITO96,SAITO97}

Our main goal is to study properties of nuclear matter and finite nuclei. 
We investigate whether the different versions of the QMC model lead to results 
which are consistent with established hadronic phenomenology.
At present such an analysis might provide the only testing ground for the
reliability of the strikingly different predictions for the changes of the 
internal nucleon structure.

A well established framework for relativistic hadronic models is provided
by quantum hadrodynamics (QHD) \cite{SEROT97}.
Numerous calculations have established that relativistic mean-field models 
based on QHD lead to a realistic description of nuclear matter and of the 
bulk properties of finite nuclei throughout the Periodic 
Table \cite{SEROT97,REINHARD86,FPW87,FURNSTAHL89,BODMER89,GAMBHIR90,%
FURNSTAHL93,FST97}.
A very compelling feature of QHD has been the reproduction of spin-orbit 
splittings in finite nuclei.
For comparison we employ
a QHD model that includes quartic and cubic scalar self-interactions.
We calibrate the model parameters so that QMCII, MQMC and QHD lead to the same
nuclear matter properties at equilibrium.
The QMCI does not provide a sufficient number of parameters to fit
all the desired nuclear matter properties. Particularly, a
rather high value for the effective nucleon mass at equilibrium is a 
prediction in this model \cite{SAITO94a}.

In our previous work \cite{MUELLER97,MUELLER98} we demonstrated that the QMC 
models are equivalent
to a QHD type mean field model with a general nonlinear scalar potential
and a coupling to the gradients of the scalar field.
One of the key observations in the success of hadronic models is
that nonlinear scalar self-interactions, {\em i.e.} a nonlinear potential,
must be included 
\cite{REINHARD86,FPW87,FURNSTAHL89,BODMER89,GAMBHIR90,FURNSTAHL93,FST97,%
FST96,BOGUTA77}.
This potential is constrained by nuclear observables and we check if the 
QMC models predict the typical size and form of the potential.

In nuclear matter we find that the models which can be accurately 
calibrated, namely QMCII, MQMC and QHD, lead to essentially identical results.
The MQMC and QMCII models reproduce large scalar and vector potentials
and the typical size of the  nonlinear potential.
In contrast, the QMCI model leads to rather small scalar and vector mean fields.
At the equilibrium point the nonlinear potential has only half the size
of the value predicted by the other models.

The results for the binding energies and spin-orbit splittings are similar.
The QMCII, MQMC and QHD model, which reproduce the empirical properties of 
nuclear matter, lead to a realistic description of the 
experimental numbers; the only exception are the light nuclei. The surface
energy in the QMCII model is too small and as a consequence the light systems
are systematically overbound.
The QMCI model underestimates the binding energies 
and gives only a poor reproduction of single-particle spectra 
and spin-orbit splittings \cite{GUICHON96,SAITO96,BLUNDEN96,SAITO97}.
The key observations is that the MQMC and QMCII model 
can reproduce sufficient small values for the effective mass, 
which is strongly correlated with the spin-orbit force in relativistic 
mean-field models.

The main difference between the models arises from the nonlinear 
coupling to the gradient of the scalar field.
The coupling has no effect in nuclear matter and its size is a prediction
of the underlying bag model. It mainly effects 
the nuclear shapes leading to more diffuse surfaces in the QMC models. 
At normal nuclear matter density the gradient coupling in the QMCII model
is 6-7 times bigger than in the MQMC model.
From the point of view of an effective field theory 
a gradient coupling arises naturally as a subset of possible
nonlinear meson-meson interactions \cite{FST97}. 
However, rather than attempting to compete with these more sophisticated
versions of QHD, it is our goal to analyze if an approach which is based 
on a simple quark model can reproduce well established results of nuclear 
phenomenology.  For comparison we employ a more conventional version of QHD
which does not include the gradient coupling.
We compensate the effect on the nuclear shapes by adjusting the scalar mass 
to reproduce the experimental value of the charge radius in $^{40}{\rm Ca}$.
Although after the adjustment all the models predict nearly identical rms 
charge radii we find sizable differences in the predicted density profiles.
The QMCI and QMCII model lead to very compact nuclei with small
central densities and steep surface areas.
This effect is more pronounced for the light nuclei.

The outline of this paper is as follows:
In Sec.~II, we review the formalism for finite nuclei and nuclear matter.
Section III contains a short summary of the QMC model and the relations which
determine the properties of the nucleon. We also briefly discuss the 
calibration procedure.
In Sec.~IV, we analyze nuclear matter. We concentrate on the predictions
for the nonlinear potential.
In Sec.~V, we apply the models to finite nuclei.
Sec.~VI contains a short summary and our conclusions.
%
%%%%%%%%%%%%%%%%%%%%%%%%%%%%%%%%%%%%%%%%%%%%%%%%%%%%%%%%%%%%%%%%%%%
%%%%%%%%%%%%%%%%%%%%%%%%%%%%%%%%%%%%%%%%%%%%%%%%%%%%%%%%%%%%%%%%%%%
%
\section{The Quark-Meson Coupling Model for Nuclear Systems}
To study the properties of finite nuclei and nuclear matter we use a 
relativistic mean-field model containing nucleons, neutral scalar ($\phi$) 
and vector fields ($V_{\mu}$) and the isovector $\rho$ meson field
($\mathbf{b}_\mu$).
For a realistic description of finite nuclei, 
the electromagnetic field ($A_{\mu}$) must also be included.
We assume the nucleons obey the Dirac equation
\begin{equation}
\left(i\not{\!\partial} 
       - \not{\!\mathcal{ V}} 
       - {1\over 2}\mbox{\boldmath$\tau \cdot$}
         \not{\!\mbox{\boldmath$\mathcal{B}$}}
       - {1\over 2}(1+\tau_3)\not{\!\!\mathcal{A}} - M^*\right) 
       \psi_{\scriptscriptstyle N} (x) = 0 \ .
\label{eq:diracn}     
\end{equation}
The potentials 
($\mathcal{V}^\nu,\mbox{\boldmath$\mathcal{ B}$}^\nu,
\mathcal{A}^\nu$)
and the effective mass $M^*$ are functionals of the meson mean fields,
their form depends on the underlying quark model. 

In the QMC model the
quarks are described by the Dirac equation
\begin{equation}
\left(i\not{\!\partial} 
       - g_{\rm v}^q \not{\!V}
      - {1\over 2} g_\rho^q \mbox{\boldmath$ \tau \cdot \not{b}$}
       - (1+3\tau_3){e\over 6} \not{\!\!A}
       - [m_q- g_{\rm s}^q \phi]\right) \psi_q (x) = 0 \ ,
\label{eq:diracq}     
\end{equation}
where $m_q$ is the current quark mass. The quark wave function is
subject to the bag model boundary conditions at the surface of the bag. 
Because quarks and nucleons interact with the meson mean fields,
Eqs.~(\ref{eq:diracn}) and (\ref{eq:diracq}) define a 
self-consistent scheme for the description of the nuclear system.
In infinite nuclear matter (${\mathcal{A}}^{\nu}=0$) the meson mean fields are
constant and the potentials are given by \cite{SAITO94a,GUICHON96,SAITO96}
\begin{eqnarray}
{\mathcal{V}}^\nu &=&  3 g_{\rm v}^q \, V^{\nu}  
               \equiv g_{\rm v} \, V^{\nu} \ , \nonumber \\ 
\mbox{\boldmath$\mathcal{B}$}^\nu &=& g_{\rho}^q \, {\mathbf{b}}^\nu
         \equiv g_{\rho} \, {\mathbf{b}}^\nu \ .
\label{eq:pots}     
\end{eqnarray}
The effective mass is an ordinary function of the scalar field, {\em i.e.}
\begin{equation}
M^* = M^* (\phi) \ .
\label{eq:potm}     
\end{equation}
For a finite system the solution of Eq.~(\ref{eq:diracn}) and
Eq.~(\ref{eq:diracq}) is rather complicated due to
the variation of the meson mean fields over the bag volume. 
In consequence, the quark wave function and the ground state of a bound nucleon
are no longer spherically symmetric \cite{BLUNDEN96}. 
To make a numerical solution 
feasible it is necessary to calculate the quark properties by using
some suitably averaged form for the meson mean fields. Here we follow the
prescription of \cite{GUICHON96,SAITO96,SAITO97} and replace the meson mean 
fields on the quark level
by their value at the center of the nucleon bag, {\em i.e.} we neglect
the spatial variation of the mean fields over the bag volume. In this local
density approximation the potentials in Eq.~(\ref{eq:diracn}) 
are simply obtained by the corresponding nuclear matter relations given by
Eqs.~(\ref{eq:pots}) and Eq.~(\ref{eq:potm}).
The corresponding relation for the electromagnetic field is
\begin{eqnarray}
{\mathcal{A}}^\nu &=& e \, A^\nu \ .
\end{eqnarray}

If we restrict considerations to spherically symmetric nuclei only the
$V_0$ component of the neutral vector field and the neutral $\rho$ meson field
(denoted by $b_0$) contribute.
The ground state energy of a nucleus can be written as
\begin{eqnarray}
E_{\scriptscriptstyle N} &=& \sum_{i=occ.} E_i 
    + \int \, dV \Biggl( [{1\over 2}(\nabla \phi)^2+ U_{\rm s}(\phi)]
    - {1\over 2}[(\nabla V_0)^2+ m_{\rm v}^2 V_0^2]  \label{eq:energyn}\\
& &\null - {1\over 2}[(\nabla b_0)^2+ m_\rho^2 b_0^2]
         - {1\over 2}(\nabla A_0)^2\Biggr)  \ , \nonumber
\end{eqnarray}
where $E_i$ are the eigenvalues of the Dirac equation Eq.~(\ref{eq:diracn}).

To analyze a general class of models, we introduce a nonlinear scalar 
potential of the form
\begin{equation}
U_{\rm s}(\phi)={1\over 2} m^2_{\rm s}\phi^2
+{1\over 3!} \kappa\phi^3
+{1\over 4!} \lambda\phi^4
   \ .        \label{eq:us}
\end{equation}
The actual mean field configuration is obtained by extremization of the 
energy. This leads to the set of self-consistency equations
\begin{eqnarray}
\Delta \phi - {\partial U_{\rm s}\over \partial \phi} &=& 
{\partial \over \partial \phi} M^*(\phi)\rho_{\rm s}
\ ,      \label{eq:eqscalar} \\
\bigl(\Delta - m_{\rm v}^2 \bigr) V_0 &=& -g_{\rm v} \rho
\ ,      \label{eq:eqvector} \\
\bigl(\Delta - m_\rho^2 \bigr) b_0 &=& -g_\rho {1\over 2} \rho_3
\ ,      \label{eq:eqiso} \\
     \Delta A_0 &=& -e \rho_p
\ .      \label{eq:eqem} 
\end{eqnarray}
The densities on the right-hand side are the nuclear densities calculated
with the wave functions in Eq.~(\ref{eq:diracn}):
\begin{eqnarray}
\rho_{\rm s} &=& \sum_{i=occ.} \bar\psi_{\scriptscriptstyle N}^i 
\psi_{\scriptscriptstyle N}^i
\ ,      \label{eq:rscalar} \\
\rho &=& \sum_{i=occ.} \bar\psi_{\scriptscriptstyle N}^i 
\gamma^0 \psi_{\scriptscriptstyle N}^i
\ ,      \label{eq:rvector} \\
\rho_3 &=& \sum_{i=occ.} \bar\psi_{\scriptscriptstyle N}^i 
\tau_3 \gamma^0 \psi_{\scriptscriptstyle N}^i
\ ,      \label{eq:riso} \\
\rho_p &=& {1\over 2}\sum_{i=occ.}\bar\psi_{\scriptscriptstyle N}^i 
(1+\tau_3) \gamma^0 \psi_{\scriptscriptstyle N}^i
\ .      \label{eq:rem} 
\end{eqnarray}
In the limit of infinite symmetric nuclear matter Eqs.~(\ref{eq:energyn}) 
simplifies to
\begin{eqnarray}
{E_{\scriptscriptstyle N}\over V} = 
  {2\over \pi^2}\int_{0}^{k_{\rm F}}dk\, k^2(k^2+M^*{}^2)^{1/2}
   + g_{\rm v} V_0\rho
   - {1\over 2} m_{\rm v}^2 V_0^2
   + U_{\rm s}(\phi)
   \ .        \label{eq:energybag}
\end{eqnarray}
For the time-like component of the vector field 
Eq.~(\ref{eq:eqvector}) reduces to
\begin{eqnarray}
g_{\rm v} V_0
= {g_{\rm v}^2 \over m_{\rm v}^2} \rho \ ,
\label{eq:vector}
\end{eqnarray}
whereas the scalar field is determined by the equivalent of 
Eq.~(\ref{eq:eqscalar}):
\begin{eqnarray}
{\partial U_{\rm s}\over \partial \phi}
=-{\partial M^* \over \partial \phi} \  {2 M^*\over \pi^2}
\int_{0}^{k_{\rm F}}dk\, {k^2\over(k^2+M^*{}^2)^{1/2}} \ .
\label{eq:scalarb}
\end{eqnarray}
In symmetric matter the Fermi momentum of the nucleons is related to the 
conserved baryon density by
\begin{eqnarray}
\rho={2\over 3\pi^2}k_{\rm F}^3 \ .
         \label{eq:baryondensity}
\end{eqnarray}
The details of the underlying quark substructure are entirely contained in 
the expression for the effective mass $M^*(\phi)$. In the next section we 
will discuss the functional form of the effective mass in the framework of 
the QMC model.
%
%%%%%%%%%%%%%%%%%%%%%%%%%%%%%%%%%%%%%%%%%%%%%%%%%%%%%%%%%%%%%%%%%%%
%%%%%%%%%%%%%%%%%%%%%%%%%%%%%%%%%%%%%%%%%%%%%%%%%%%%%%%%%%%%%%%%%%%
%
\section{The Quark-Meson Coupling Model}
In this section, we briefly summarize the relations which determine the 
nuclear equation of state in the quark-meson coupling model. 
For further details we refer the reader to Refs.~\cite{SAITO94a,JIN96a,JIN96b}.

In the QMC model the nucleon in the nuclear medium is described as a
static, spherical MIT bag in which quarks couple to meson mean fields.

The energy of a bag consisting of three quarks in the ground state can 
be expressed as
\begin{eqnarray}
E_{bag} = 3 {\Omega_q\over R}-{Z\over R}+{4\over 3}\pi R^3 B  \ . 
 \label{eq:ebag}
\end{eqnarray}
where the parameter $Z$ accounts for the zero point motion and $B$ is the
bag constant.
The coupling of the quarks to the scalar field is inherent in the quantities
$\Omega_q$ and $x$ which are given by
\begin{eqnarray}
\Omega_q &=& \sqrt{x^2+(R m_q^*)^2} \nonumber\\
j_0(x)&=&\left({\Omega_q-R m_q^*\over \Omega_q+R m_q^*}\right)^{1/2}j_1(x) \ ,
\label{eq:bessel}
\end{eqnarray}
and where $m_q^* = m_q^0-g_{\phi}^q\phi$ denotes the effective 
quark mass and $m_q^0$ is the current quark mass. In the following
we choose $m_q^0=5$ MeV.

To remove the spurious center-of-mass motion in the bag we follow
Ref.~\cite{GUICHON96} and adjust the parameter $Z$ in Eq.~(\ref{eq:ebag})
to reproduce the experimental value of the nucleon mass in the vacuum,
{\em i.e.} we take
\begin{equation}
M^*(\phi) = E_{bag} \ . \label{eq:massdef}
\end{equation}

For a fixed meson mean-field configuration
the bag radius $R$ is determined by the equilibrium
condition for the nucleon bag in the medium
\begin{eqnarray}
{\partial M^* \over \partial R} = 0 \ . \label{eq:equilibrium}
\end{eqnarray}
In free space $M$ can be fixed
at its experimental value 939 MeV and the condition 
Eq.~(\ref{eq:equilibrium}) to determine the parameters $B=B_0$ and $Z=Z_0$.
For our choice, $R_0=0.8$ fm, the result for $B_0^{1/4}$ and $Z_0$ are
169.97 MeV and 3.295, respectively. 

In the original version of the QMC model\cite{GUICHON88,SAITO94a}
the bag parameters $B$ and $Z$
were held fixed at their free space values $B=B_0, Z=Z_0$.
Formally, the bag constant $B$ is associated with the QCD trace anomaly.
In the nuclear environment it is expected to decrease with increasing density
as argued in Ref.~\cite{ADAMI93}.

To account for this physics in the QMC approach 
Jin and Jennings \cite{JIN96a,JIN96b} proposed two models for the
medium modification of the bag constant:
a direct coupling model in which the bag constant is a function of the
scalar field and a scaling model which relates the bag constant directly
to the effective nucleon mass.
The density dependence is then generated self-consistently in terms of 
these in-medium quantities.
In a previous work \cite{MUELLER97} this approach was generalized and 
it was demonstrated that the resulting improved MQMC model can be 
accurately calibrated to reproduce the empirical properties of
nuclear matter \cite{MUELLER97} and finite nuclei \cite{MUELLER98}.

For our purpose here we adopt the model ${\rm MQMC}_{\rm A}$ of 
Ref.~\cite{MUELLER97} in which the
bag constant depends on the scalar field only
\begin{eqnarray}
{B\over B_0}=\left(1-g_{\scriptscriptstyle\rm B}
{\phi\over M}F(\phi)\right)^{\eta}
\quad \hbox{\rm with} \quad F(0)=1 \ .
\label{eq:ansatzdc}
\end{eqnarray}
We model the functional form of $F$ by using a simple polynomial 
parametrization
\begin{eqnarray}
F(\phi)=1+\alpha\phi+\beta\phi^2 \ .
\label{eq:poly}
\end{eqnarray}
The functional form Eq.~(\ref{eq:poly}) provides sufficient flexibility
and it is not necessary to include further nonlinearities in the scalar
potential. We therefore analyze the MQMC model with
\begin{equation}
U_{\rm s}(\phi)={1\over 2} m^2_{\rm s}\phi^2
   \ .        \label{eq:usmqmc}
\end{equation}
In addition we consider the original QMC model in two versions.
In the first model, which we denote QMCI, we disregard the nonlinear terms
in the scalar potential and use the same form as in Eq.~(\ref{eq:usmqmc}).
In the second model, which we denote QMCII, we include the cubic and quartic 
terms as given by Eq.~(\ref{eq:us}). We analyze if these additional 
nonlinearities lead to an improvement of the model.

We close this section with a brief description of the calibration procedure.
The parameters $B_0$ and $Z$ are fixed to reproduce the nucleon mass in
the vacuum.
In nuclear matter the MQMC model contains seven free parameters. 
The parametrization of the bag constant contains the parameter $\eta$ and
the three couplings $(g_{\scriptscriptstyle\rm B}, \alpha, \beta)$; 
in addition values for the ratios
$g_{\rm s}^q/m_{\rm s}, g_{\rm v}/m_{\rm v}, g_\rho/m_\rho$ 
are needed. 

To determine the parameters we follow \cite{MUELLER97}. 
For given values
of $\eta$ and $g_{\rm s}^q/m_{\rm s}$ we determine the values of
$(g_{\scriptscriptstyle\rm B}, \alpha, \beta, g_{\rm v}/m_{\rm v}, 
g_\rho/m_\rho)$ to reproduce the
equilibrium properties of nuclear matter, which are taken to be:
the equilibrium density and binding energy 
($\rho^{\scriptscriptstyle 0}$,
$-e_{\scriptscriptstyle 0}$),
the nucleon effective mass at equilibrium 
($M_{\scriptscriptstyle 0}^*$) 
the symmetry energy ($a_{\scriptscriptstyle 4}$) 
and the compression modulus ($K_{\scriptscriptstyle 0}$).
The set of equilibrium properties used here are listed
in the first row of Table~\ref{tab:one} (denoted by input). 
For more details concerning the calibration procedure we refer the reader to 
Ref.~\cite{MUELLER97}.

The model QMCII contains the five parameters
$g_{\rm s}^q/m_{\rm s}, g_{\rm v}/m_{\rm v}, g_\rho/m_\rho, \kappa$ 
and $\lambda$ 
which are determined to reproduce the same equilibrium properties as for the 
MQMC model.
In the model QMCI only the parameters
$g_{\rm s}^q/m_{\rm s}, g_{\rm v}/m_{\rm v}, g_\rho/m_\rho$ are available. 
We chose to fix these three parameters such that the model reproduces the same 
density, binding energy and symmetry energy as in MQMC and QMCII. 
Correspondingly, the effective mass and compression modulus at equilibrium are 
a prediction. The values are quoted in the second row of Table~\ref{tab:one}. 
We find $M_{\scriptscriptstyle 0}^*/M=0.80$ and 
$K_{\scriptscriptstyle 0}= 280$ MeV.

For finite nuclei calculations values for the meson masses are needed. 
The mass of the scalar meson $m_{\rm s}$ is determined to reproduce the
charge radius in $^{40}{\rm Ca}$ as we will discuss in more detail in
section V.  The masses of the remaining mesons are fixed at their experimental
values $m_{\rm v} = 783$  MeV and $m_\rho = 770$ MeV.
%
%%%%%%%%%%%%%%%%%%%%%%%%%%%%%%%%%%%%%%%%%%%%%%%%%%%%%%%%%%%%%%%%%%%
%%%%%%%%%%%%%%%%%%%%%%%%%%%%%%%%%%%%%%%%%%%%%%%%%%%%%%%%%%%%%%%%%%%
%
\section{Properties Of Nuclear Matter}
Quark-meson coupling models are designed to describe both bulk properties
of nuclear systems and medium modifications of the internal structure of
the nucleon.
It is important that the models reproduce established results of nuclear
phenomenology before reliable predictions for changes of the quark 
substructure can be made.

\begin{figure}[tbhp]
\centerline{%
\vbox to 6in{\vss
   \hbox to 3.3in{\includegraphics{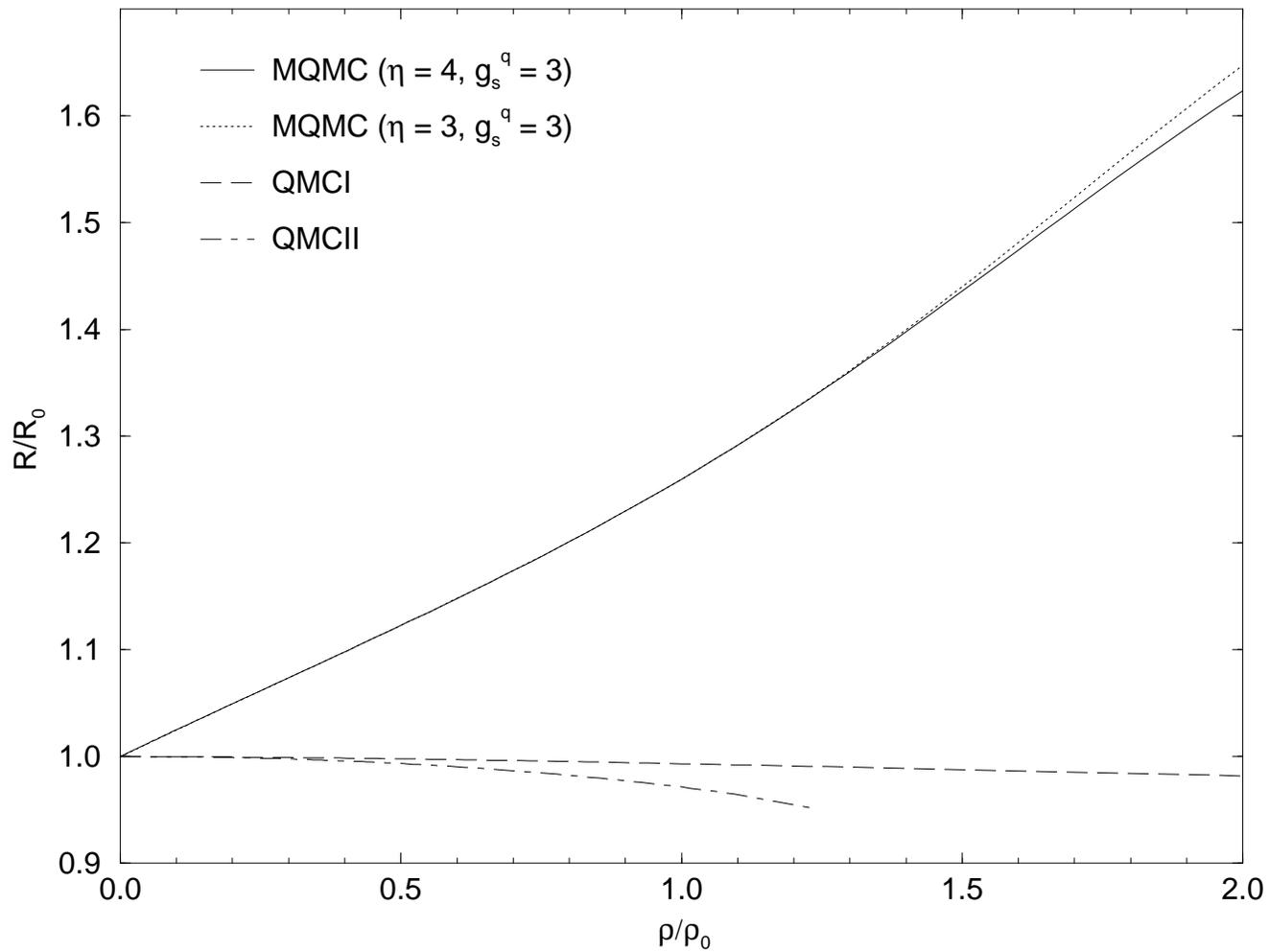}\hss}}
}
\caption{Bag radius as a function of the density for the models
MQMC, QMCI and QMCII.
The solution for the QMCII model terminates at 
$\rho_{\scriptscriptstyle\rm max}\approx 1.23 \rho^{\scriptscriptstyle 0}$.}
\label{fig:radius}
\end{figure}

An important quantity is the nucleon size which is represented
in the framework of the QMC model by the bag radius \cite{JIN97}.
The density dependence of the bag radius can be seen in 
Fig.~\ref{fig:radius}. 
The opposite behavior of the prediction of the models is striking.
The MQMC model leads to significantly swollen 
nucleons \cite{JIN96a,JIN96b,MUELLER97}. 
At equilibrium the bag radius increases to roughly
$25\%$ of its free space value. At low and moderate densities we observe 
a very small dependence on the model parameters. In contrast, 
for the QMC models the bag radius decreases slightly with
increasing density \cite{SAITO94a,GUICHON96,SAITO96,SAITO97}. 
The effect is more pronounced for the model QMCII.

\begin{figure}[tbhp]
\centerline{%
\vbox to 6in{\vss
   \hbox to 3.3in{\includegraphics{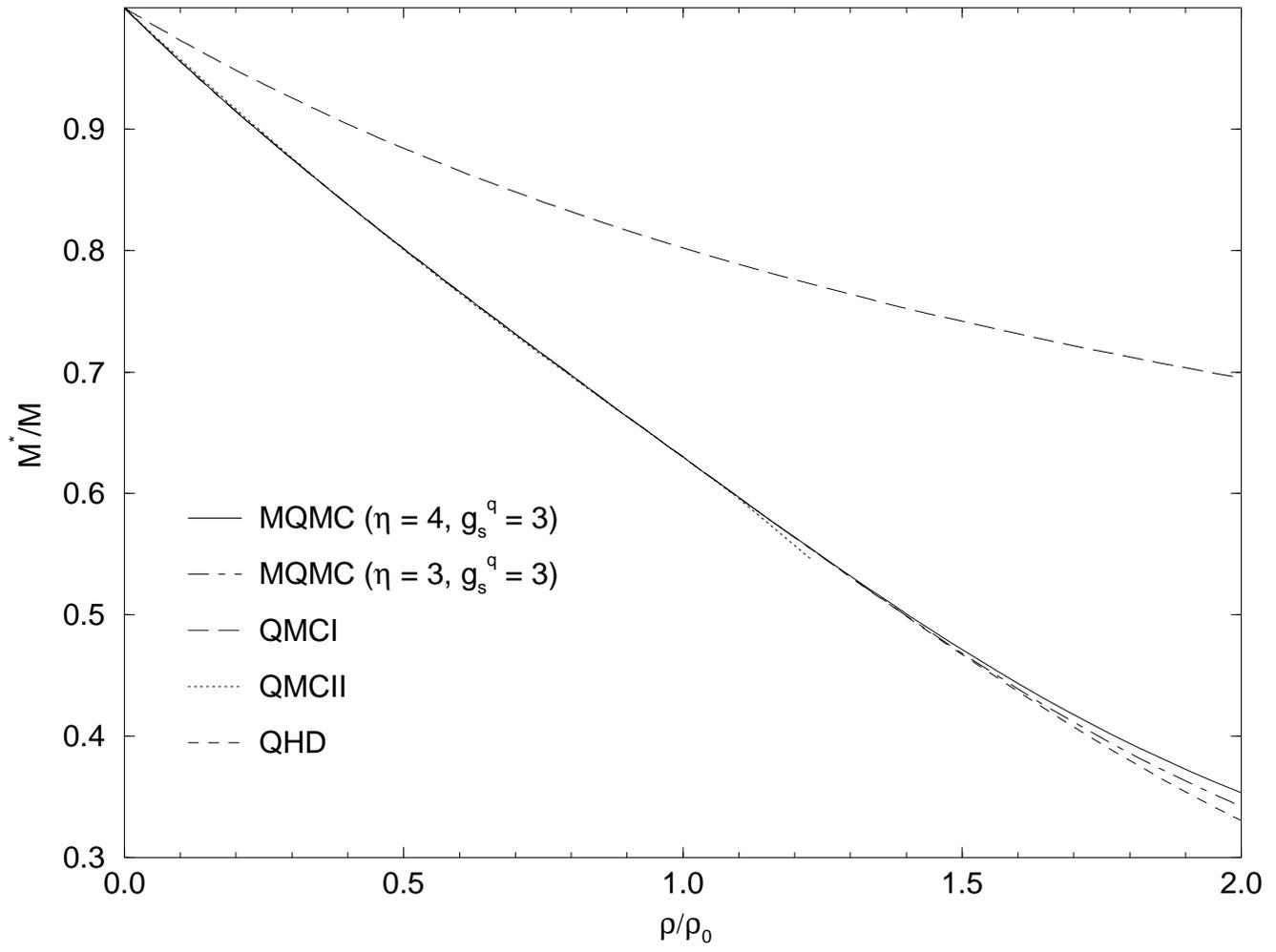}\hss}}
}
\caption{Effective nucleon mass as a function of the density.}
\label{fig:mstar}
\end{figure}

The medium dependence of the effective nucleon mass is of central
importance in relativistic nuclear phenomenology.
The effective mass is shown in Fig.~\ref{fig:mstar} as a function of
the density.
Also indicated is the QHD result based on a model which includes a nonlinear
scalar potential of the form given by Eq.~(\ref{eq:us}). The QHD parameters are 
determined by reproducing the equilibrium properties listed in 
Table~\ref{tab:one}.
The masses for MQMC, QMCII and QHD are nearly identical at low and moderate
densities. In contrast, QMCI predicts a very high and slowly decreasing 
effective mass \cite{SAITO94a}.

The bag models cannot be extrapolated to arbitrary high densities. The solutions
cease to exist when the point $x=0$ with $R m_q^*=-3/2$ in 
Eq.~(\ref{eq:bessel}) is reached.
This corresponds to a maximal density 
$\rho_{\scriptscriptstyle\rm max}\approx 4.92 \rho^{\scriptscriptstyle 0}$
and
$\rho_{\scriptscriptstyle\rm max}\approx 1.23 \rho^{\scriptscriptstyle 0}$
for the QMCI and QMCII models respectively.
The rather small value for the QMCII model severely limits its applicability.
The MQMC model leads to
$\rho_{\scriptscriptstyle\rm max}\approx 3.38 \rho^{\scriptscriptstyle 0}$.
Here applications are limited by the large bag radii rather than by the
maximal density. The individual bags start overlapping before 
$\rho_{\scriptscriptstyle\rm max}$ is reached \cite{MUELLER97}.

\begin{figure}[tbhp]
\centerline{%
\vbox to 6in{\vss
   \hbox to 3.3in{\includegraphics{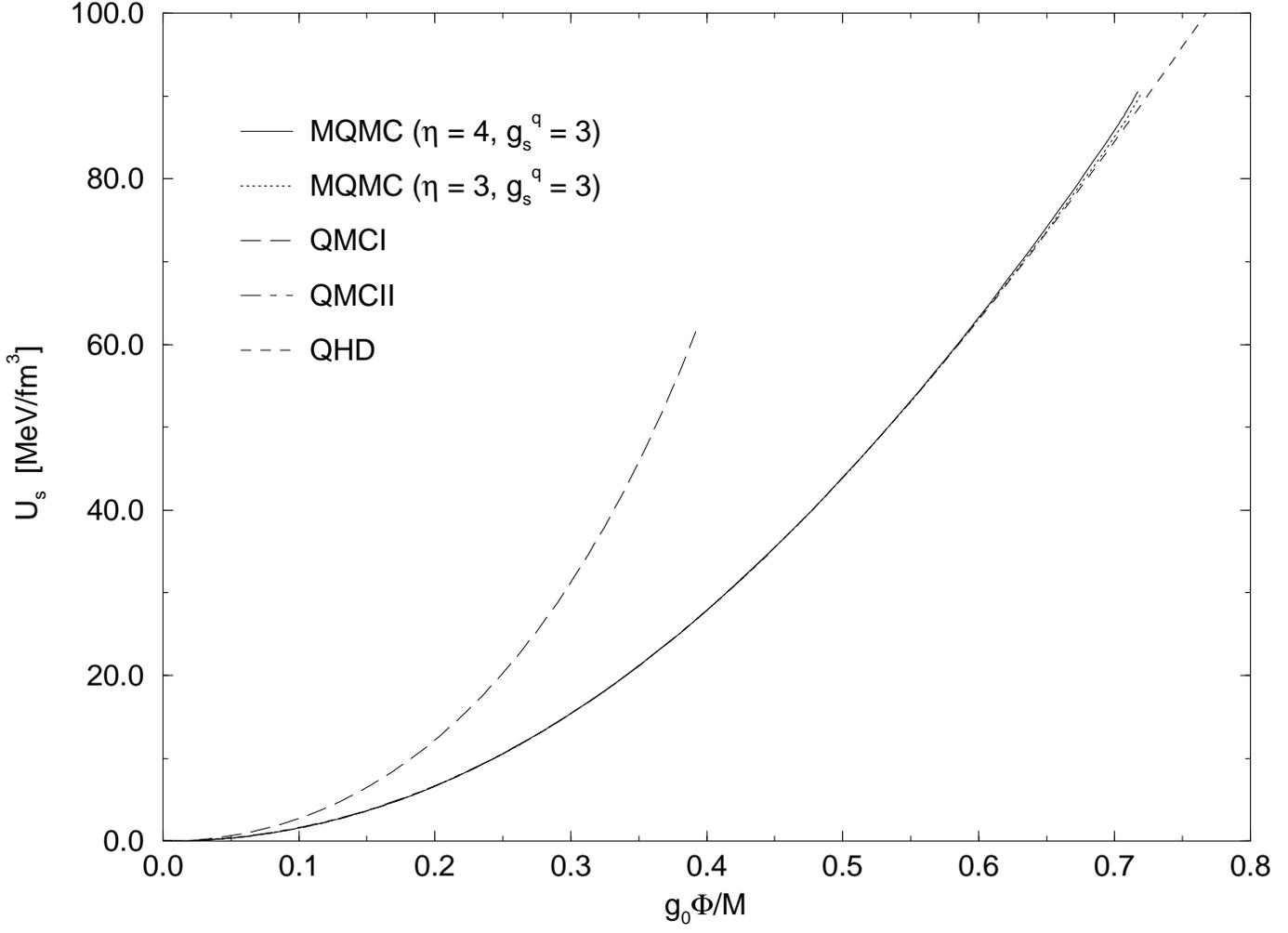}\hss}}
}
\caption{
Predicted nonlinear scalar potential as a function of the transformed
scalar field $g_{\scriptscriptstyle 0} \Phi=M-M^*$.}
\label{fig:pot}
\end{figure}

As discussed in Ref.~\cite{MUELLER97,MUELLER98} there is a direct
relation between the QMC models and QHD-type mean field models. 
The main difference between QMC and QHD is the functional form of the
effective mass. In QMC it is a complicated function of the scalar field
\begin{equation}
M^*_{\scriptscriptstyle\rm QMC}= M^*_{\scriptscriptstyle\rm QMC}(\phi) \ ,
\end{equation}
whereas in QHD it is linearly related to the scalar field
\begin{equation}
M^*_{\scriptscriptstyle\rm QHD}= M- g_{\scriptscriptstyle 0} \Phi \ .
\end{equation}
This suggests a redefinition of the scalar field in QMC \cite{MUELLER97}:
\begin{eqnarray}
g_{\scriptscriptstyle 0} \Phi (\phi) 
\equiv M-M_{\scriptscriptstyle\rm QMC}^*(\phi) 
 =  M- \Bigr( 3 {\Omega_q\over R}-{Z\over R}+{4\over 3}\pi R^3 B  \Bigl)
\ . \label{eq:trans}
\end{eqnarray}
The coupling $g_{\scriptscriptstyle 0}$
is chosen to normalize the new field according to
\begin{eqnarray}
\Phi (\phi) {\mathop{=}_{\phi \to 0}} \phi + O(\phi^2) \ ,
\nonumber
\end{eqnarray}
and is given by
\begin{eqnarray}
g_{\scriptscriptstyle 0} = 
- {\partial M^* (\phi)\over \partial \phi }\biggr|_{\phi=0} \ .
\label{eq:geff}
\end{eqnarray}
The scalar potential in Eq.~(\ref{eq:us}) 
can now be expressed in terms of the new field
\begin{eqnarray}
U_{\rm s}(\Phi) \equiv
U_{\rm s}\bigl(\phi(\Phi)\bigr)=
{1\over 2} m_{\rm s}^2 \phi^2(\Phi)+{1\over 3!} \kappa\phi^3(\Phi)
+{1\over 4!} \lambda\phi^4(\Phi) \ .
\label{eq:transpot}
\end{eqnarray}
The field redefinition Eq.~(\ref{eq:trans}) recasts the QMC model into
a QHD type mean field model. Most importantly, it predicts a specific form
for the nonlinear scalar potential.
Fig.~\ref{fig:pot} indicates the predicted potentials as a function of the
transformed scalar field given by Eq.~(\ref{eq:trans}).
Below the saturation point ($g_{\scriptscriptstyle 0}\Phi/M=0.37$) 
the curves for MQMC and QMCII are
almost identical to the QHD result. 
For a given value of the scalar field the QMCI potential is considerably 
bigger. At the saturation point ($g_{\scriptscriptstyle 0}\Phi/M= 0.2$),
however, it is only half the size of the other models.

\begin{figure}[tbhp]
\centerline{%
\vbox to 5in{\vss
   \hbox to 3.3in{\includegraphics{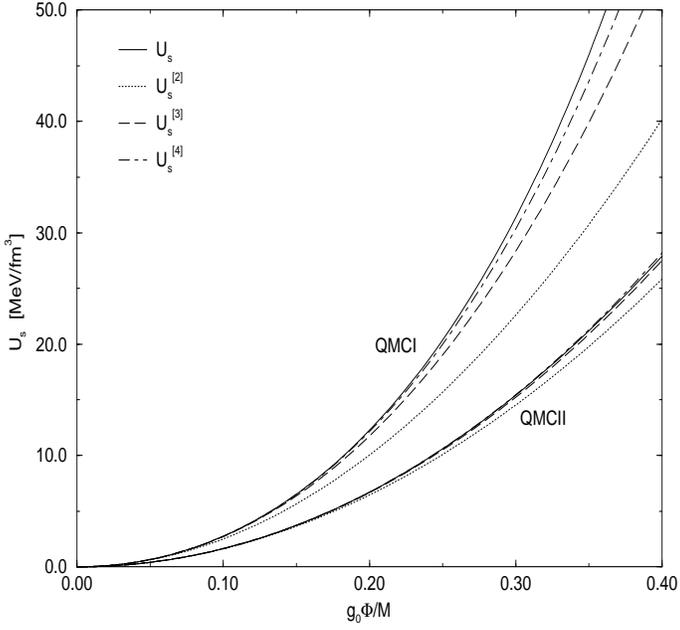}
                  \includegraphics{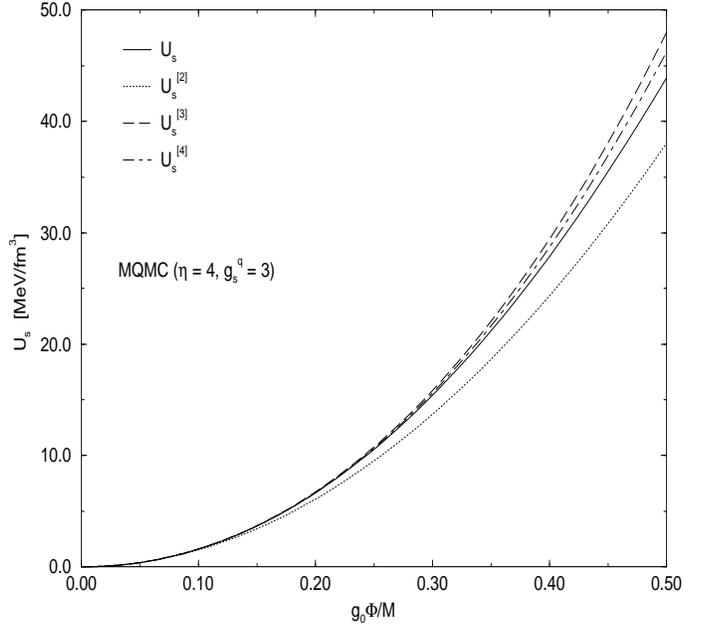}\hss}}
}
\centerline{\vbox to 0.3in{}}
\centerline{(a) \hbox to 3.5in{} (b)}
\centerline{\vbox to 0.5in{}}
\caption{Second, third and fourth order contributions to the nonlinear
scalar potential. In part (a) we consider the models QMCI and QMCII,
and part (b) indicates the result for the MQMC model.} 
\label{fig:polypot}
\end{figure}

The agreement in Fig.~\ref{fig:pot} is somewhat misleading. 
Although there is good agreement between QHD, MQMC and QMCII
the functional form of the potentials is different.
This can be studied by expanding the potential in Eq.~(\ref{eq:transpot})
in terms of the scalar field
\begin{eqnarray}
U_{\rm s}\bigl(\phi(\Phi)\bigr)=
{1\over 2} \varepsilon_2 (g_0 \Phi)^2
+{1\over 3!} \varepsilon_3(g_0 \Phi)^3
+{1\over 4!} \varepsilon_4(g_0 \Phi)^4 + \ \ldots \ .
\label{eq:potex}
\end{eqnarray}
Fig.~\ref{fig:polypot} indicates $U_{\rm s}$ and the series in 
Eq.~(\ref{eq:potex}) truncated at second, third and fourth order. 
Part (a) shows the result for QMCI and QMCII.
The potential is well represented by the fourth order polynomial 
in Eq.~(\ref{eq:potex}). 
The major contribution comes from the quadratic term.
The very small deviation of the truncated fourth order series from the exact 
potential indicates that higher order corrections are negligible.
At the saturation point we find that the neglected higher order 
terms give a correction of only $2\%$.
The result for the MQMC model is indicated in part (b) of 
Fig.~\ref{fig:polypot}.
Here the third and fourth order contributions are more important than
for the QMC models.
At the equilibrium point the neglected higher order contribution
give a correction of roughly $3\%$.

It is well known that nonlinear self-interactions of the form 
Eq.~(\ref{eq:us}), or Eq.~(\ref{eq:potex}), must be included for a 
successful low-energy nuclear phenomenology
\cite{REINHARD86,FPW87,FURNSTAHL89,BODMER89,GAMBHIR90,FURNSTAHL93,FST97,%
FST96,BOGUTA77}.
Our analysis demonstrates that once the QMCII and MQMC models are calibrated
by using characteristic properties of nuclear matter they predict the typical
size of these nonlinear self-interactions.
Furthermore, all the models appear to be {\em natural}, meaning that the 
nonlinear potential is well represented by a low order 
polynomial\footnote{For a more complete discussion of naturalness in the 
QMC model, see Ref.~\cite{SAITO97b}.}.

The details of the underlying quark structure are entirely contained in the 
coefficients of the series in Eq.~(\ref{eq:transpot}).
Based on a nuclear matter analysis alone, where the MQMC and the QMCII model
produce equivalent results, it is not possible to decide which
prediction for the changes of the internal nucleon structure is more
reliable.
%
%%%%%%%%%%%%%%%%%%%%%%%%%%%%%%%%%%%%%%%%%%%%%%%%%%%%%%%%%%%%%%%%%%%
%%%%%%%%%%%%%%%%%%%%%%%%%%%%%%%%%%%%%%%%%%%%%%%%%%%%%%%%%%%%%%%%%%%
%
\section{Consequences For Finite Nuclei}
The field redefinition in Eq.~(\ref{eq:trans}) can also be applied in a 
finite system.
Expressed in terms of the new field $\Phi$
the contribution of the scalar field to the energy in Eq.~(\ref{eq:energyn})
is given by
\begin{eqnarray}
E_{\rm s} &=& \int \, dV [{1\over 2}(\nabla \phi)^2+ U_{\rm s}(\phi)]
\label{eq:energys} \\
    &=&  \int \, dV [{1\over 2}(\nabla \Phi)^2 H(\Phi)^2 + U_{\rm s}(\Phi)] \ ,
\nonumber
\end{eqnarray}
In addition to the nonlinear potential Eq.~(\ref{eq:transpot})
the transformation also induces a coupling to the gradients of the scalar field
\begin{eqnarray}
H(\Phi)=1+h(\Phi)={\partial \phi\over \partial \Phi} 
       = -{g_{\scriptscriptstyle 0}\over {\partial M^*\over\partial \phi}} \ .
\label{eq:hbag}
\end{eqnarray}
The coupling $h(\Phi)$ has no effect in nuclear matter calculations and
is a prediction of the underlying bag model.
From a modern point of view, our model contains a subset of possible nonlinear
meson-meson couplings.
In more sophisticated versions of QHD \cite{FST97},
inspired by concepts and methods of effective field theory, 
these terms and many others are considered. 
For comparison we employ a conventional version of QHD
which contains the standard form of the nonlinear potential
given by Eq.~(\ref{eq:us}) and with $h(\Phi)=0$.

\begin{figure}[tbhp]
\centerline{%
\vbox to 6in{\vss
   \hbox to 3.3in{\includegraphics{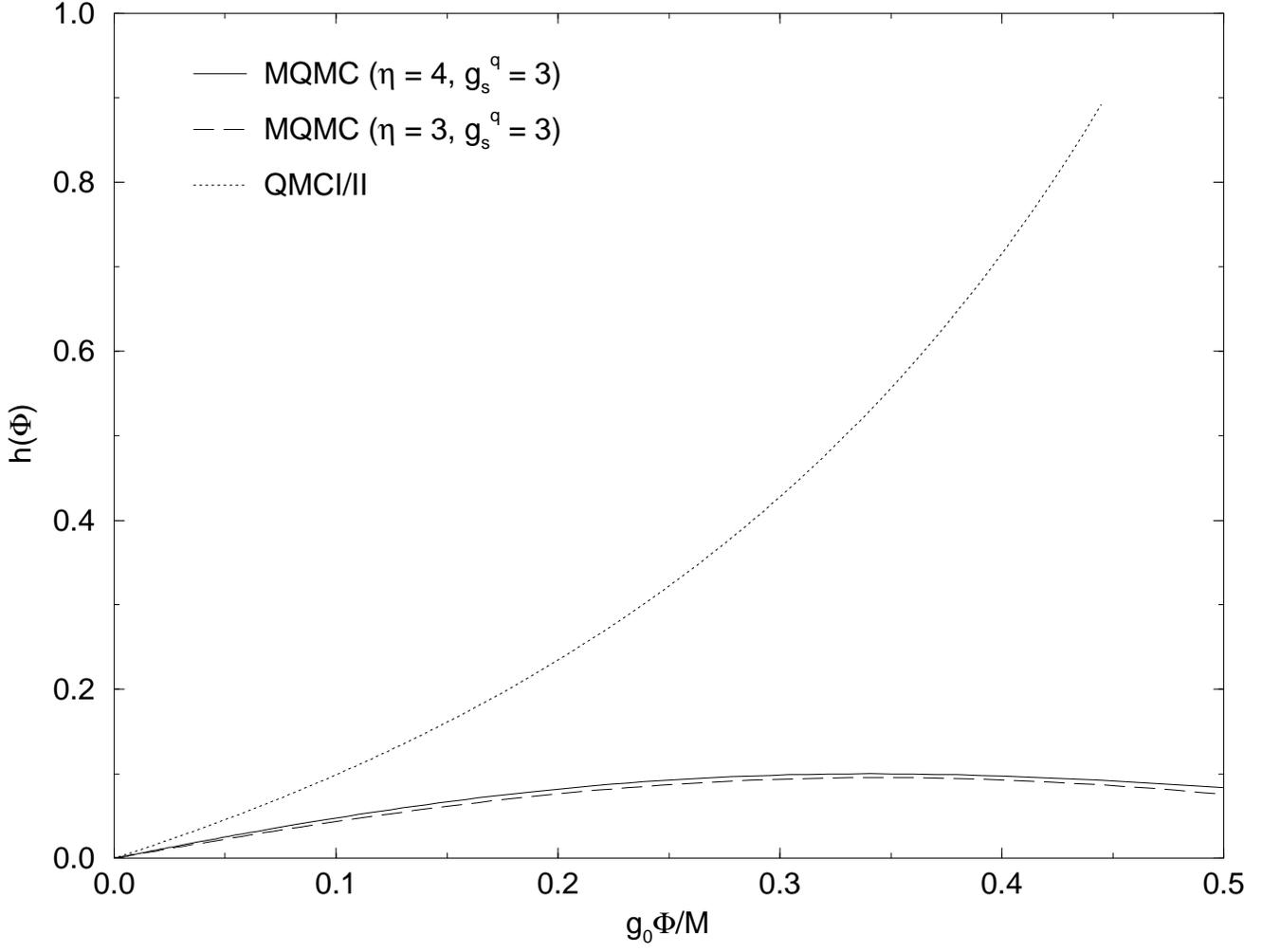}\hss}}
}
\caption{Gradient coupling as a function of the transformed
scalar field $g_{\scriptscriptstyle 0} \Phi=M-M^*$.
The function $h(\Phi)$ is identical for the models QMCI and QMCII.}
\label{fig:grad}
\end{figure}

The coupling $h(\Phi)$ is indicated in Fig.~\ref{fig:grad}. 
Relevant for applications to finite
nuclei is the region below $g_{\scriptscriptstyle 0}\Phi/M=0.37$ for QMCII
and MQMC and below $g_{\scriptscriptstyle 0}\Phi/M=0.2$ for QMCI
which corresponds to the saturation point of nuclear matter. 
The MQMC model predicts very small values for the gradient coupling.
The function $h(\Phi)$ is identical for the two versions of the
QMC model. Here the coupling decreases significantly. Near nuclear
matter equilibrium the coupling is 6 to 7 times bigger than for the MQMC
model.

\begin{figure}[tbhp]
\centerline{%
\vbox to 5in{\vss
   \hbox to 3.3in{\includegraphics{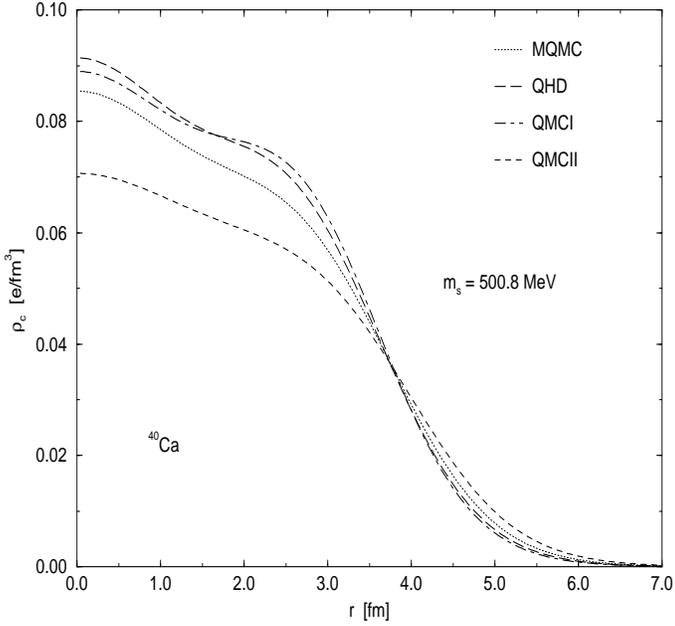}
                  \includegraphics{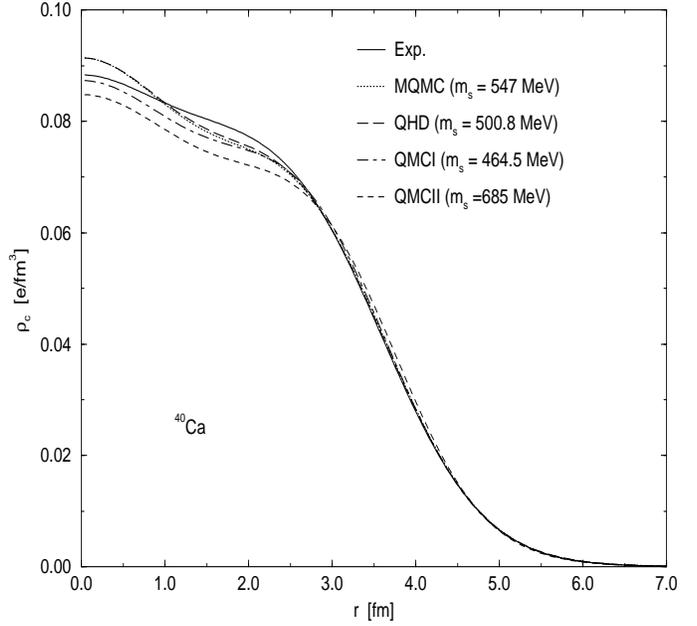}\hss}}
}
\centerline{\vbox to 0.3in{}}
\centerline{(a) \hbox to 3.5in{} (b)}
\centerline{\vbox to 0.5in{}}
\caption{Charge density for $^{40}{\rm Ca}$.
In part (a) the same mass for the scalar meson ($m_{\rm s}=500.8$MeV) was used 
for all the models. 
In part (b) the scalar mass was adjusted to
reproduce the experimental charge radius in $^{40}{\rm Ca}$.
The parameters for the MQMC model are $\eta=4$ and $g_s^q = 3$.}
\label{fig:rcca}
\end{figure}

The gradient coupling $h(\Phi)$ has an impact on the bulk and single particle 
properties of a nucleus. For positive values it effectively decreases the
coupling strength of the scalar density to the scalar field $\Phi$ leading 
to significantly smaller mean fields.
The most prominent effect are changes of the nuclear surface. 
This can be studied in Fig.~\ref{fig:rcca} which indicates the charge density 
for $^{40}{\rm Ca}$. Charge densities and charge radii are calculated by 
convoluting the point proton density Eq.~(\ref{eq:rem}) 
with an empirical proton charge form factor \cite{HOROWITZ81}.
In part (a) the same mass for the scalar meson ($m_{\rm s}=500.8$MeV) was used 
for all the models. Note that all the models, except QMCI, predict the same 
equilibrium 
properties of nuclear matter. The gradient coupling changes the surface
drastically. The density becomes smaller in the interior region and the surface
area is more diffuse. As can be expected from Fig.~\ref{fig:grad} the effect is 
more pronounced for QMCII. To compensate for this effect we adjust the 
scalar mass
such that the models reproduce the experimental charge radius in 
$^{40}{\rm Ca}$.
The resulting charge densities are indicated in part
(b) of Fig.~\ref{fig:rcca}. Also included is the experimental charge
density \cite{VRIES87}.
The adjusted values for the scalar mass are an indication of the size of
the gradient coupling. 
The masses had to be increased by 10 and $30\%$ for the MQMC and QMCII models
respectively.
After the adjustment the curves for QHD and MQMC are almost identical. 
Due to the very large value of the scalar mass the QMCII model leads to very 
compact nuclear shapes.
The density is still too small in the interior region and the surface area is 
now very steep.
Also included in Fig.~\ref{fig:rcca} is the result for QMCI. 
Here the mean fields are much smaller than in the other models 
and the impact of the gradient coupling is 
weaker than for QMCII. In order to reproduce the charge 
radius of $^{40}{\rm Ca}$ the scalar mass has to be decreased.

\begin{figure}[tbhp]
\centerline{%
\vbox to 6in{\vss
   \hbox to 3.3in{\includegraphics{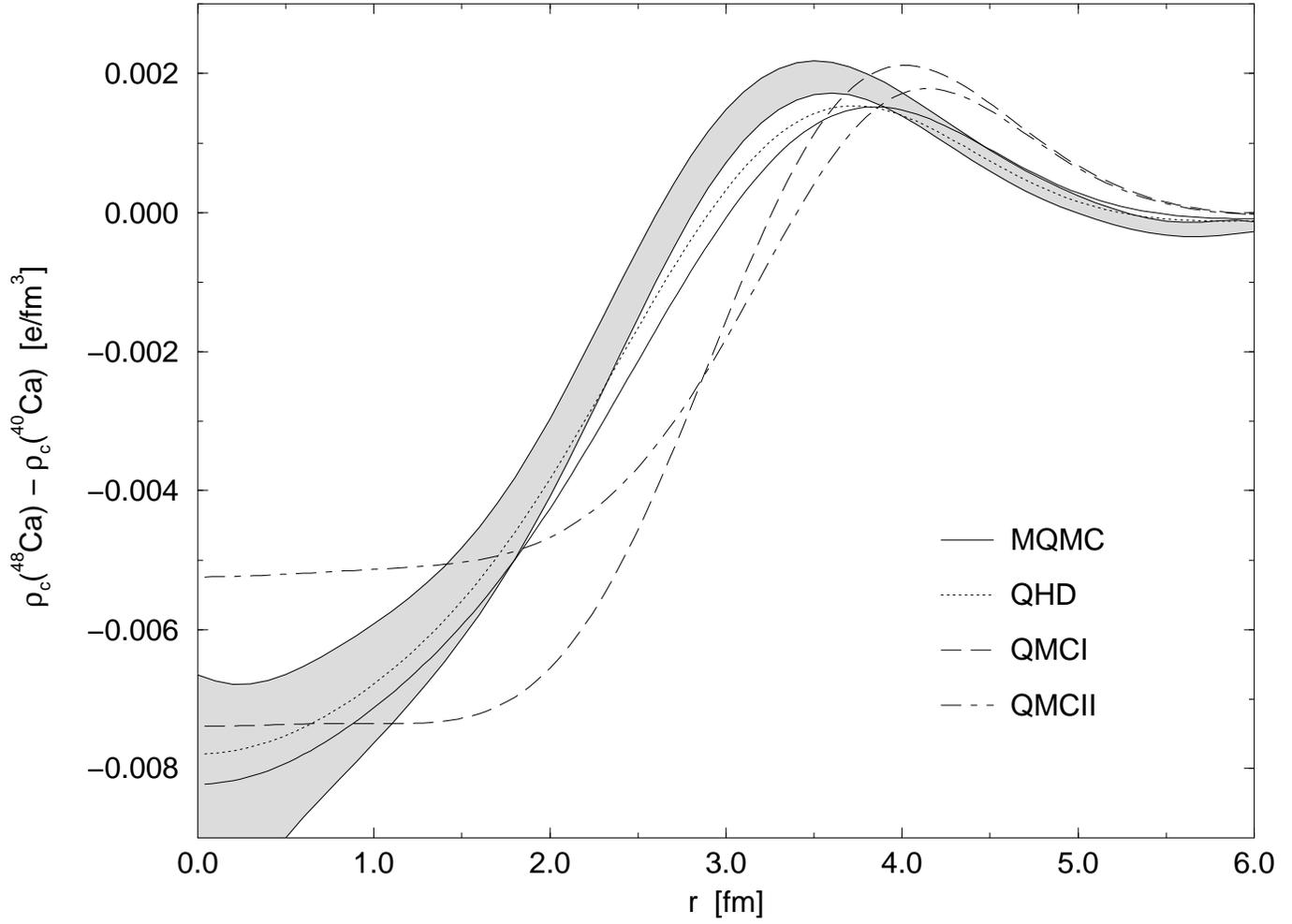}\hss}}
}
\caption{Isotope shift in calcium. The difference between the
$^{48}{\rm Ca}$ and $^{40}{\rm Ca}$ charge densities is shown.
The experimental (shaded) curve is taken from Ref.~\protect\cite{FROIS79}.
The parameters for the MQMC model are $\eta=4$ and $g_s^q = 3$.}
\label{fig:drcca}
\end{figure}

\begin{figure}[tbhp]
\centerline{%
\vbox to 6in{\vss
   \hbox to 3.3in{\includegraphics{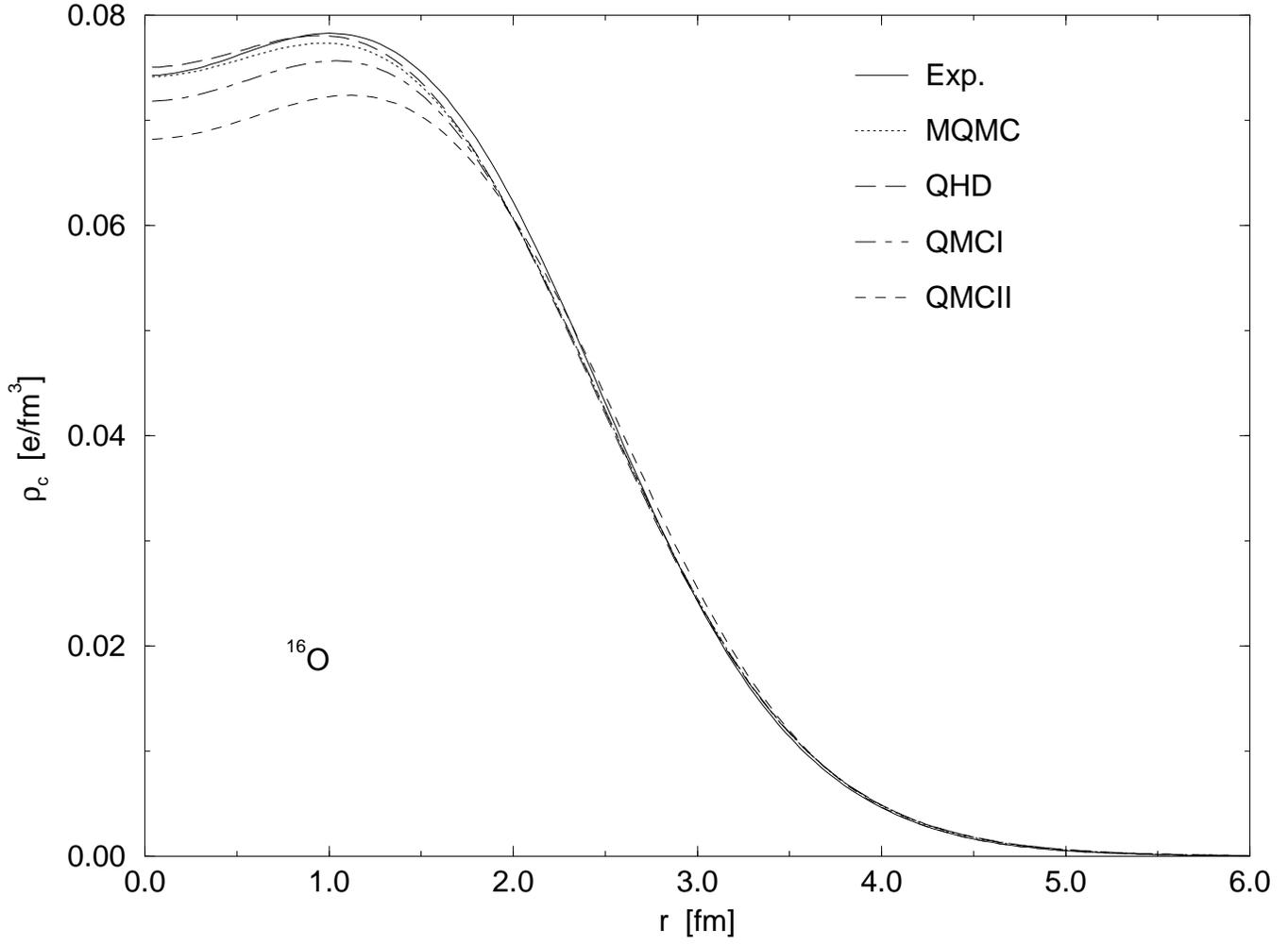}\hss}}
}
\caption{Charge density for $^{16}{\rm O}$.
The parameters for the MQMC model are $\eta=4$ and $g_s^q = 3$.}
\label{fig:rco}
\end{figure}

\begin{figure}[tbhp]
\centerline{%
\vbox to 6in{\vss
   \hbox to 3.3in{\includegraphics{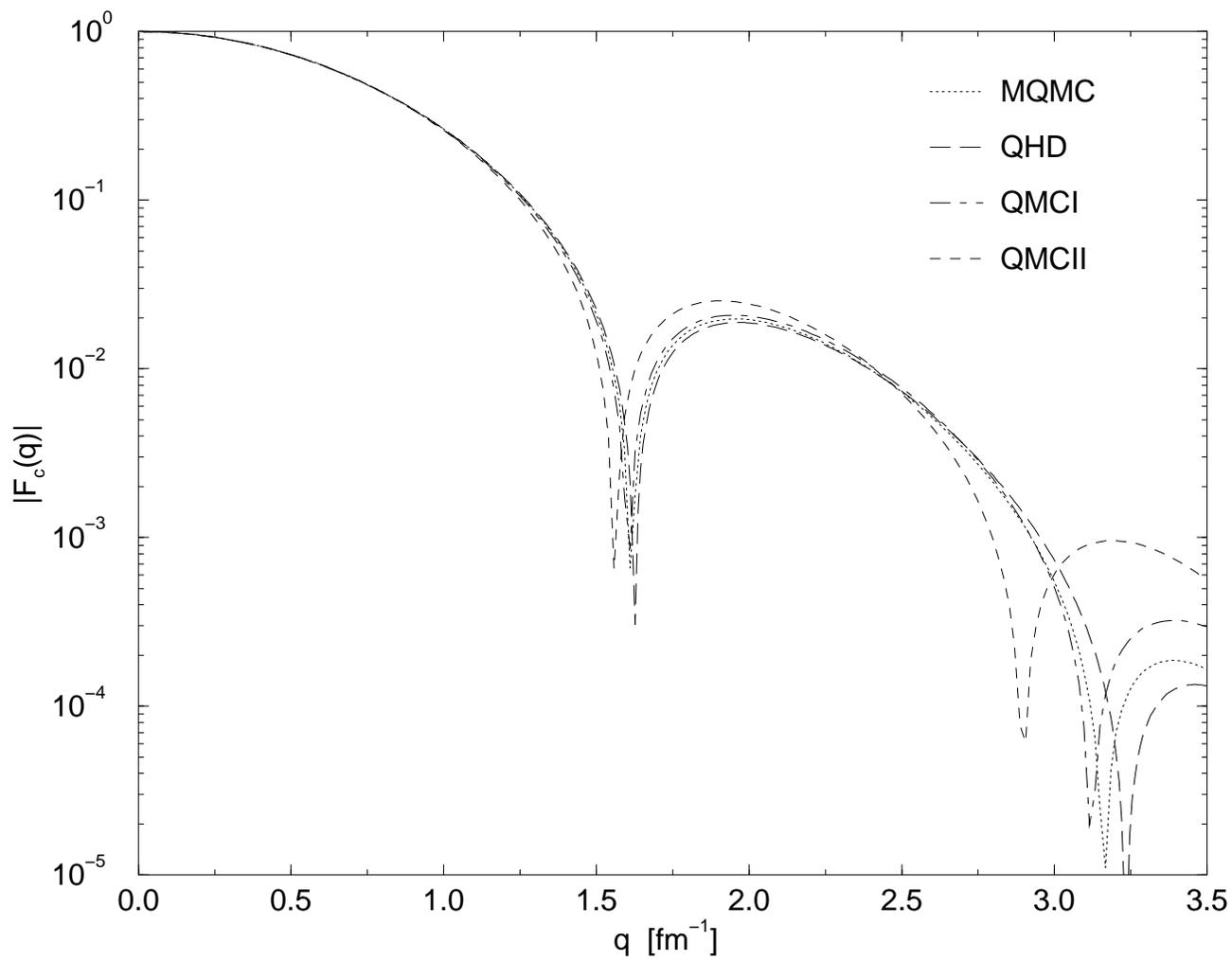}\hss}}
}
\caption{Form factor for $^{16}{\rm O}$.
The parameters for the MQMC model are $\eta=4$ and $g_s^q = 3$.}
\label{fig:fco}
\end{figure}

The effect of the gradient coupling on the nuclear shapes depends
strongly on the mass number of the system.
Discrepancies in the predictions of the models are much smaller for the 
heavier nuclei.
The quality of the reproduced nuclear shapes for the lighter nuclei
can be studied in Figs.~\ref{fig:drcca}-\ref{fig:fco}.
Fig.~\ref{fig:drcca} indicates the isotope shift in calcium. Shown is the
difference of the $^{48}{\rm Ca}$ and $^{40}{\rm Ca}$ charge density. 
The QMCI and QMCII model cannot provide a realistic description of the
experimental curve \cite{FROIS79}.
In Fig.~\ref{fig:rco} we show the charge density of $^{16}{\rm O}$. 
Similarly in Fig.~\ref{fig:rcca} the QMCI and II models 
underestimate the density in the interior region. The corresponding
form factor is indicated in Fig.~\ref{fig:fco}. For the diffraction radius,
which indicates the location of the first zero of the form factor, the 
predictions are in reasonable agreement. At larger momenta a substantial model 
dependence in the diffraction pattern arises. 

Binding energies and rms charge radii for $^{16}{\rm O}, ^{40}{\rm Ca},
^{48}{\rm Ca}, ^{90}{\rm Zr}$ and $^{208}{\rm Pb}$ are shown 
in Table~\ref{tab:two}
for various values of $\eta$ and $g_{\rm s}^q$. Also included are the QHD,
QMCI, and QMCII results and the experimental values.
To make a more realistic comparison with the experimental data the last
row of Table~\ref{tab:two} indicates c.m. corrections taken from 
Ref.~\cite{NEGELE70}.
Overall the models QMCII, MQMC and QHD give a realistic description of the 
binding energies and radii. Including the c.m. corrections the QHD and MQMC
model reproduce the experimental binding energies within an accuracy of $3\%$. 
For MQMC we observe a small model dependence. Due to changes in the surface 
systematic the binding energies increase with $\eta$ and $g_{\rm s}^q$ 
\cite{MUELLER98}. The rms charge radii are insensitive to the parametrization.
The QMCII model systematically overestimates the binding energies of the 
light nuclei. The error is maximal for Oxygen ($\sim 6\%$) and decreases with 
increasing mass number. We attribute this overbinding to the small surface 
energy. 
The binding energies in the QMCI model are too small.
The error of the theoretical predictions ranges between $6\%$ and $9\%$.

For QMCI we could not reproduce the results for the binding energies given in
Ref.~\cite{SAITO96} (see their Table~ 4), 
particularly for the light nuclei. We believe that the 
discrepancy is due to their approximate treatment of the nucleon mass. 
The authors of Ref.~\cite{SAITO96} employed a simple parametrization of the 
form
\begin{eqnarray}
M^*(\phi)\approx M+{d M^*\over d\phi}\Biggr|_{\phi=0}\ \phi+c\phi^2 \ .
\label{eq:fit}
\end{eqnarray}
Although the parametrization is quite accurate we found that the binding
energies are sensitive to small variations of the fit parameter $c$. 

The impact of the gradient coupling is less drastic for the binding energies 
than for the nuclear shapes. To obtain a quantitative estimate 
of the surface energy we followed Ref.~\cite{HOROWITZ81} and fitted the 
error of the calculated energies $\delta e_{\scriptscriptstyle N}$ 
(including the c.m. correction) to a form
\begin{eqnarray}
\delta e_{\scriptscriptstyle N} = \alpha + {\beta\over A^{1/3}} .
\nonumber
\end{eqnarray}
The coefficient $\beta$ which indicates the correction to the surface
energy is negative for all the models. We found $\beta= -2.7$ MeV for
the QMCII model, $\beta =(-1)-(-1.5)$ MeV for the MQMC models and 
$\beta = -0.8$ for the QHD model. Although the effect is small, the
QMCII model clearly produces the smallest surface energies.

\begin{figure}[tbhp]
\centerline{%
\vbox to 6in{\vss
   \hbox to 3.3in{\includegraphics{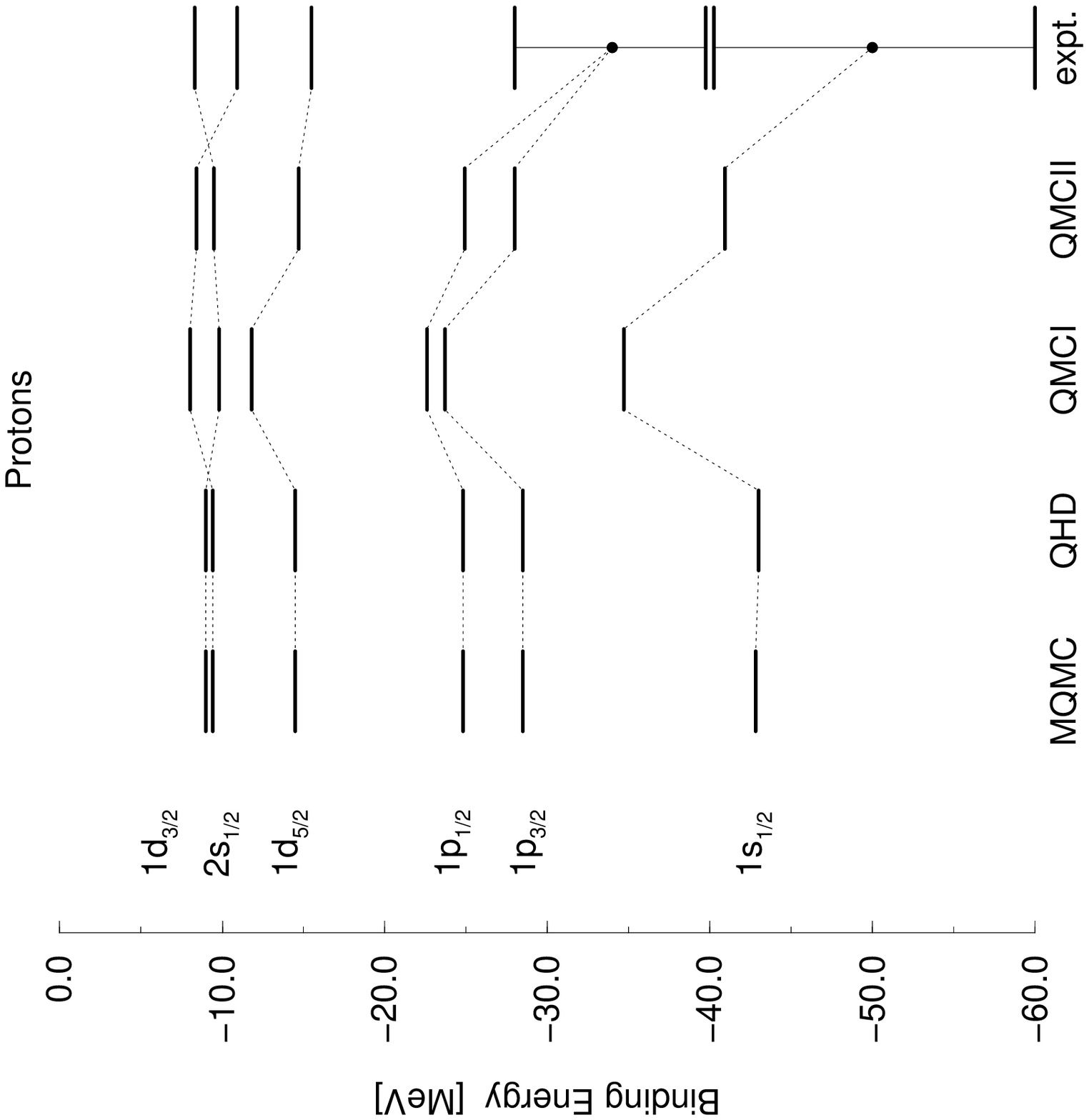}
                  \includegraphics{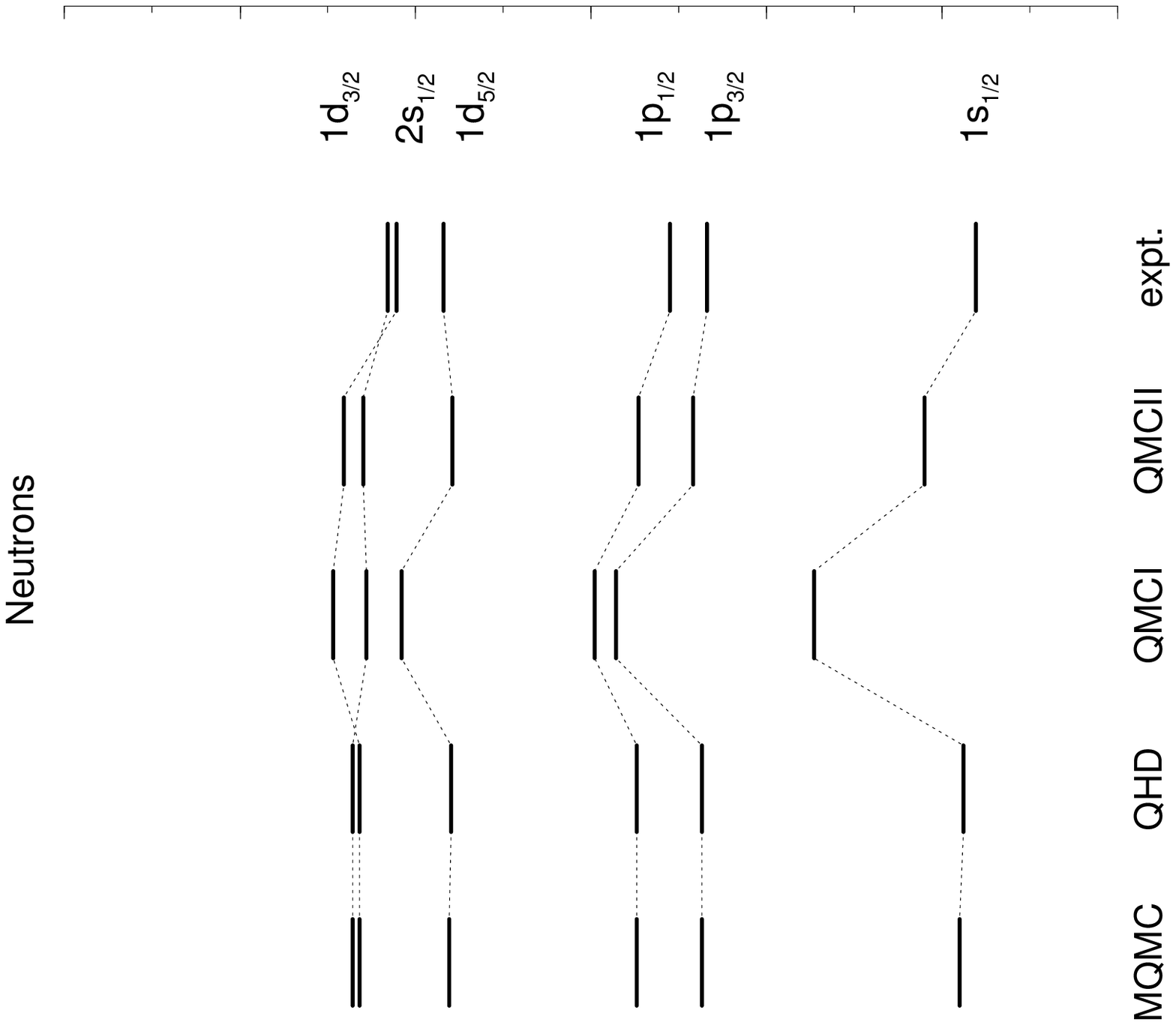}\hss}}
}
\caption{Single particle spectrum of $^{40}{\rm Ca}$.
The parameters for the MQMC model are $\eta=4$ and $g_s^q = 3$.
The experimental numbers are taken from Ref.~\protect\cite{CAMPI72}.}
\label{fig:spectrum}
\end{figure}

As stated in earlier references~\cite{SAITO96,BLUNDEN96,SAITO97} the QMCI model 
leads to a fair
description of the bulk properties of nuclei but gives only a poor
reproduction of single-particle spectra and spin-orbit splittings.
Spin-orbit splittings are highly correlated to the effective nucleon
mass which is too high in the model QMCI.
In view of these shortcomings the analysis of single-particle spectra and 
spin-orbit splittings is very important.
The single-particle levels for $^{40}{\rm Ca}$ are shown in 
Fig.~\ref{fig:spectrum}.
The MQMC model and QHD clearly give a more realistic description of the energy 
levels than the two versions of QMC.
The energies of the deeply bound states are too small in QMCI.
Both QMCI and QMCII predict an incorrect level ordering of the $2s_{1/2}$ and 
$1d_{3/2}$ states (see also Refs.~\cite{SAITO96,SAITO97}).
We observe a very good agreement between MQMC and QHD.
Generally, there is only a very weak dependence of the energy levels on the 
parameters $\eta$ and $g_s^q$ \cite{MUELLER98}.

Results for other nuclei are similar.
Spin-orbit splittings for the highest occupied proton and neutron states in 
$^{208}{\rm Pb}$ and $^{16}{\rm O}$ are shown in Table~\ref{tab:three}
and Table~\ref{tab:four} respectively.
The results demonstrate directly the importance of the effective nucleon
mass.
The QMCI model systematically underpredicts the splittings \cite{SAITO96}.
The small value of the effective mass in the MQMC  and QMCII model
significantly corrects this shortcoming.

\begin{figure}[tbhp]
\centerline{%
\vbox to 6in{\vss
   \hbox to 3.3in{\includegraphics{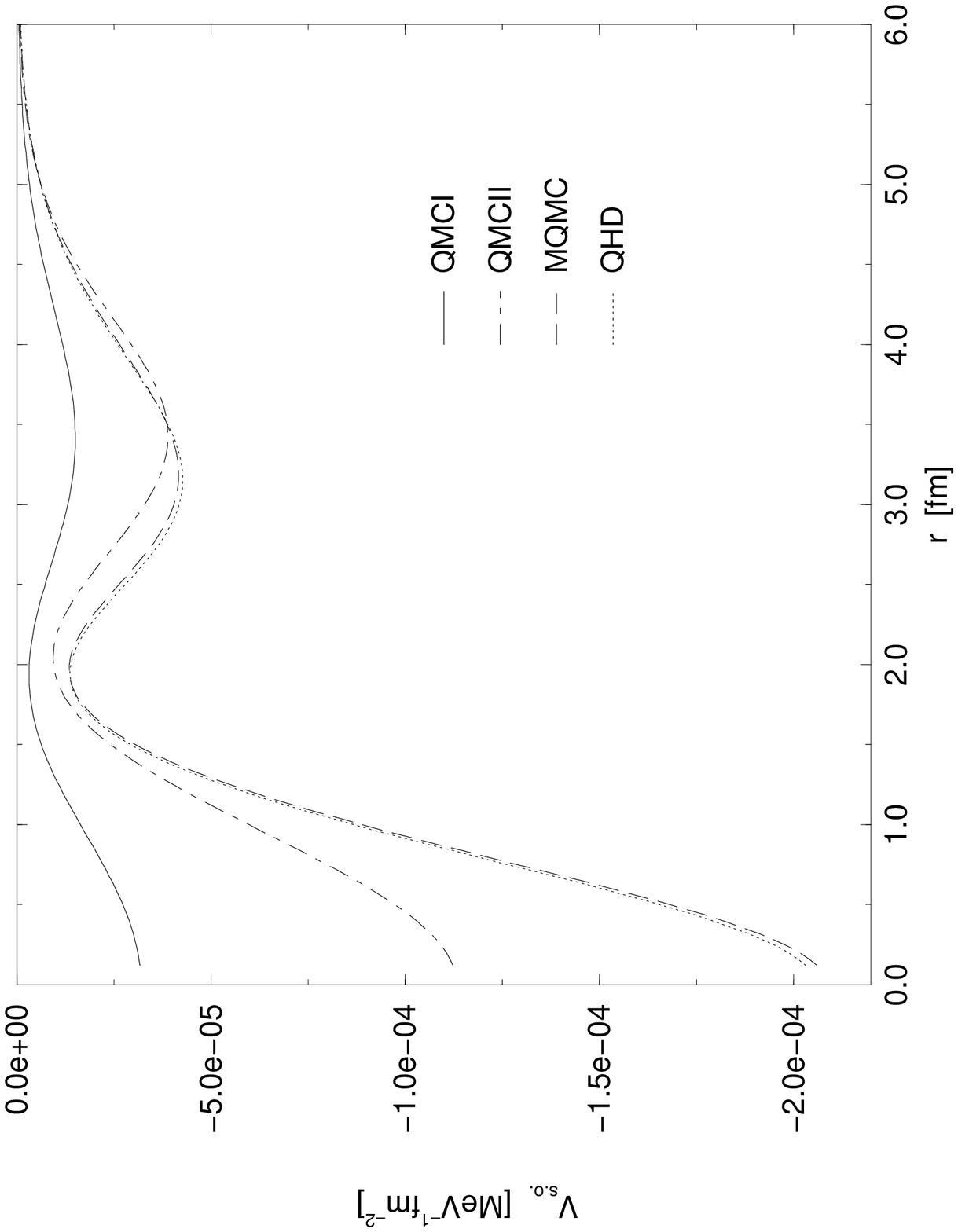}\hss}}
}
\caption{Spin-orbit potential for $^{40}{\rm Ca}$.
The parameters for the MQMC model are $\eta=4$ and $g_s^q = 3$.}
\label{fig:sopot}
\end{figure}

The spin-orbit potential for $^{40}{\rm Ca}$ can be studied in 
Fig.~\ref{fig:sopot}. It arises
in a single particle hamiltonian that acts on two-component wave functions.
Here we follow Reinhard \cite{REINHARD89} who has proposed an expansion in 
terms of a small nucleon velocity which converges better than the usual 
Foldy-Wouthuysen reduction. In this framework the spin-orbit part is given by
\begin{eqnarray}
h_{s.o.}=V_{s.o.}(r)\ \mbox{$\mathbf{\sigma\ \cdot \ L}$}
=\Biggl[{1\over \bar M^2}{1\over r}
\Biggl(g_0{d\Phi\over dr}+g_{\rm v} {dV_0\over dr}\Biggr)\Biggr]
\ \mbox{$\mathbf{\sigma\ \cdot \ L}$} \ ,
\label{eq:sopot}
\end{eqnarray}
where $\bar M$ is defined as
\begin{eqnarray}
{\bar M}=M-{1\over 2} \Bigl( g_0 \Phi + g_{\rm v} V_0 \Bigr) \ .
\end{eqnarray}
The spin-orbit potential of QMCI is rather weak explaining the small
splittings in Table~\ref{tab:three} and \ref{tab:four}. The results for QHD
and MQMC are almost identical. For QMCII the size of the potential is 
similar at the surface
of the nucleus but it has only half the strength at the origin.

The main results of our analysis are summarized in Table~\ref{tab:five}.
The column headings denote the various models discussed in this work.
The rows contain specific model features and representative results.
The first row indicates the form of the bag constant which is a function
of the scalar field in the model MQMC $(B=B(\phi))$ and which is held
at its free space value for the model QMCI and QMCII $(B=B_0)$.
The second row contains the ratio of the predicted bag radius at nuclear matter
equilibrium to its free space value.
The specific form of the scalar potential is indicated in the third row 
followed by the equilibrium value of the effective nucleon mass.

The last three rows indicate how well the models reproduce the experimental 
data. Included is the range of error for the predicted binding energies 
(including c.m. corrections). 
As a typical representative for the quality of the spin-orbit splittings
the error of the splitting of the 
$1d_{5/2}$ and $1d_{3/2}$ neutron states in $^{40}{\rm Ca}$ is shown.
The last row indicates the integrated error of the charge density in
$^{16}{\rm O}$.
%
%%%%%%%%%%%%%%%%%%%%%%%%%%%%%%%%%%%%%%%%%%%%%%%%%%%%%%%%%%%%%%%%%%%
%%%%%%%%%%%%%%%%%%%%%%%%%%%%%%%%%%%%%%%%%%%%%%%%%%%%%%%%%%%%%%%%%%%
%
\section{Summary}
In this paper we study properties of nuclear matter and finite nuclei based 
on three versions of the quark-meson coupling model.
This model describes nucleons as nonoverlapping MIT bags interacting
through scalar and vector mean fields.
The two versions QMCI and II differ from the third version denoted by MQMC 
in the treatment of the bag constant.
In the QMCI and QMCII model the bag constant is held fixed at its free space
value whereas in the MQMC model we assume it depends on the density of the
nuclear environment.
We employ a model for the bag constant in which the density dependence is
parametrized in terms of the scalar mean field.
The model QMCII is an extension of QMCI which includes additional cubic
and quartic self-interactions for the scalar meson.

The QMCII and MQMC model give rise to a sufficient number of model parameters 
which can be fit to properties of nuclear matter near equilibrium that are 
known to be characteristic of the observed bulk and single-particle properties 
of nuclei. Due to the small number of parameters not all the desired nuclear 
matter properties can be fit in the QMCI model. Particularly, the 
very high value for effective nucleon mass at equilibrium is a prediction in 
this model.

In the framework of the QMC model the nucleon size is represented by the
bag radius. Uncertainties in the predictions for the
nucleon size arise from the incomplete knowledge of the medium dependence
of the bag model parameters. The most important quantity here is the 
bag constant. 
In the MQMC model an increasing bag constant leads to significantly swollen 
nucleons in the nuclear environment. 
In contrast, for the QMCI and II model the bag constant is held fixed
and the bag radius decreases slightly with increasing density.
In light of the strikingly different predictions for the
nucleon size, the question arises whether the different models 
are all consistent with established hadronic phenomenology.
Our basic goal is to study properties of nuclear matter and finite nuclei
as an important test for the reliability of
predictions for the changes of the internal nucleon structure.

A direct relation between nuclear phenomenology and the quark
picture arises from a redefinition of the scalar field.
By introducing a new scalar field the QMC models can be cast into
a QHD-type hadronic mean-field
model with a general nonlinear scalar potential and a nonlinear coupling to
the gradients of the scalar field. 

To make contact with the established hadronic framework we compare the QMC
models with a QHD model including quartic and cubic scalar 
self-interactions calibrated to produce the same equilibrium properties of 
nuclear matter.

Our basic result is that the models QMCII, MQMC and QHD, {\em i.e.} the models
which can be calibrated to reproduce the empirical properties of nuclear
matter, lead to essentially identical results in nuclear matter and for the 
binding energies, rms charge radii and spin-orbit splittings of finite nuclei.
The MQMC and QMCII model reproduce large scalar and 
vector potentials and the typical size of the nonlinear potential.
For the binding energies and spin-orbit splittings QMCII, MQMC
and QHD lead to a realistic description of the experimental numbers.
The only exception are the lightest nuclei which are systematically
overbound in the QMCII model due to the small surface energy.
The results are in accordance with numerous calculations in the hadronic 
framework. Experience has shown that an accurate reproduction of
the nuclear matter properties leads to realistic results when the calculations
are extended to finite nuclei
\cite{REINHARD86,FPW87,GAMBHIR90,FURNSTAHL93,FST97,FST96}.

In contrast, the QMCI model does not reproduce all the empirical nuclear matter
properties. 
The scalar and vector mean fields are rather small and, as a consequence,
the binding energies are systematically underestimated.
The most prominent shortcoming of the model is
the poor reproduction of single-particle spectra 
and spin-orbit splittings \cite{GUICHON96,SAITO96,BLUNDEN96,SAITO97}.
The crucial point is that this model cannot reproduce sufficient small values 
for the effective mass, which is strongly correlated with the spin-orbit force 
in relativistic mean-field models.

The main difference between the models arises from the nonlinear 
coupling to the gradient of the scalar field.
Such a coupling is not included in our version of QHD. It effects 
the surface energy and the nuclear shape leading to a more diffuse surface 
in the QMC models. 
The size of this coupling is a pure prediction which depends on the details
of the underlying bag model.
At normal nuclear matter density the gradient coupling is 6-7 times
bigger for the QMCII model than for the MQMC model. 
The impact of the coupling is partly compensated by adjusting the scalar mass 
to reproduce the experimental value of the charge radius in $^{40}{\rm Ca}$.
Although after the adjustment all the models predict nearly identical rms 
charge radii we find that the large value of the gradient coupling in the
QMCII model leads to unrealistic features for the nuclear shapes.

These results lead us to two basic conclusions. First, it is clear that
to keep the models on a tractable level quark models for nuclear matter and 
finite nuclei always contain a number of parameters which cannot be
determined on the fundamental level. Realistic predictions for the properties
of finite nuclei require an accurate calibration of these parameters by
fitting to properties of nuclear matter near equilibrium that are
known to be characteristic of the observed bulk and single-particle properties
of nuclei. This guarantees the reproduction of large scalar and vector
potentials and a realistic description of spin-orbit splittings.
Second, the QMC models are invoked to describe medium modifications for
the internal structure of the nucleon. 
Unless more information on the fundamental level is available,
the reliability of predictions and differences which arise in different 
models must be tested by analyzing nuclear matter and finite nuclei.
Our results indicate shortcomings of the models which predict a
decreasing size of the nucleon in the nuclear environment.
The QMCI model cannot reproduce spin-orbit splittings on a satisfactory level.
Although the QMCII model corrects this shortcoming, unrealistic features
of the nuclear shapes arise.
In principle, the quality of the predicted nuclear shapes can be improved by
including additional gradient couplings. This, however, leads to new 
parameters which are more difficult to calibrate \cite{FST97}. 
As a consequence, the 
formalism looses much of its predictive power and its simplicity.
The MQMC model on the other hand, which leads to a significant increase of 
the nucleon size, provides a realistic description of nuclear matter and 
finite nuclei. At normal nuclear densities the concept of a density
dependent bag constant is very useful. However, problems arise in an 
extrapolation to higher densities. Due to the sizable increase of the bag 
radius the individual bags start overlapping and the simple bag model is no 
longer applicable.

\acknowledgements

We thank Helena H. M. Salda\~{n}a for useful comments.
This work was supported by the Natural Science and Engineering Research Council
of Canada.
%
%%%%%%%%%%%%%%%%%%%%%%%%%%%%%%%%%%%%%%%%%%%%%%%%%%%%%%%%%%%%%%%%%%%
%%%%%%%%%%%%%%%%%%%%%%%%%%%%%%%%%%%%%%%%%%%%%%%%%%%%%%%%%%%%%%%%%%%
%

%
%%%%%%%%%%%%%%%%%%%%%%%%%%%%%%%%%%%%%%%%%%%%%%%%%%%%%%%%%%%%%%%%%%%
%%%%%%%%%%%%%%%%%%%%%%%%%%%%%%%%%%%%%%%%%%%%%%%%%%%%%%%%%%%%%%%%%%%
%
\begin{table}[tbhp]
\caption{Equilibrium Properties of Nuclear Matter.
The input row denotes the nuclear matter properties which are used
to calibrate the models MQMC, QMCII and QHD. The second row indicates the
nuclear matter properties of the QMC model which contains only three free
parameters.}
\medskip
\begin{tabular}[b]{c|ccccccc}
& $(k_{\scriptscriptstyle\rm F})^{\scriptscriptstyle 0}$ 
& $\rho^{\scriptscriptstyle 0}$ 
& $M_{\scriptscriptstyle 0}^*/M$ &
  $e_{\scriptscriptstyle 0}$ & $K_{\scriptscriptstyle 0}$ & 
  $a_{\scriptscriptstyle 4}$ \\
\hline
Input &1.3\,fm$^{-1}$ & 0.1484\,fm$^{-3}$ & 0.63 & $-15.75$\,MeV & 224.2\,MeV &
35 \,MeV \\
\hline
QMCI &1.3\,fm$^{-1}$ & 0.1484\,fm$^{-3}$ & 0.80 & $-15.75$\,MeV & 280.0\,MeV &
35 \,MeV \\
\end{tabular}
\label{tab:one}
\end{table}
%
%%%%%%%%%%%%%%%%%%%%%%%%%%%%%%%%%%%%%%%%%%%%%%%%%%%%%%%%%%%%%%%%%%%
%%%%%%%%%%%%%%%%%%%%%%%%%%%%%%%%%%%%%%%%%%%%%%%%%%%%%%%%%%%%%%%%%%%
%
\begin{table}[tbhp]
\caption{Binding energy per nucleon $e_{\scriptscriptstyle N}$ (in MeV)
and rms charge radius $r_{\scriptscriptstyle c}$ in (fm) for several 
closed shell nuclei. 
}
\medskip
\begin{tabular}[b]{cc@{\extracolsep{2pt}}ccccccccccc}
\multicolumn{2}{c}{Model} & \multicolumn{1}{c}{$m_{\rm s}$} & 
                \multicolumn{2}{c}{$^{16}{\rm O}$} &
                \multicolumn{2}{c}{$^{40}{\rm Ca}$} &
                \multicolumn{2}{c}{$^{48}{\rm Ca}$} &
                \multicolumn{2}{c}{$^{90}{\rm Zr}$} &
                \multicolumn{2}{c}{$^{208}{\rm Pb}$} \\
 \cline{4-5} \cline{6-7} \cline{8-9} \cline{10-11} \cline{12-13} 
$\eta$  & $g_{\rm s}^q$ & (MeV) & 
         $e_{\scriptscriptstyle N}$ & $r_{\scriptscriptstyle c}$ &
         $e_{\scriptscriptstyle N}$ & $r_{\scriptscriptstyle c}$ &
         $e_{\scriptscriptstyle N}$ & $r_{\scriptscriptstyle c}$ &
         $e_{\scriptscriptstyle N}$ & $r_{\scriptscriptstyle c}$ &
         $e_{\scriptscriptstyle N}$ & $r_{\scriptscriptstyle c}$ \\
\hline
$3$ & $2   $ & 543 &   -7.26  & 2.74  & -8.15  & 3.48 & -8.22 & 3.50 &
                       -8.35  & 4.29   & -7.57  & 5.56  \\
$3$ & $3   $ & 544 &   -7.28  & 2.74  & -8.20  & 3.48 & -8.27 & 3.50 &
                       -8.38  & 4.29   & -7.59  & 5.56  \\
$4$ & $2   $ & 546 &   -7.37  & 2.74  & -8.26  & 3.48 & -8.33 & 3.50 &
                       -8.42  & 4.29   & -7.63  & 5.56  \\
$4$ & $3   $ & 547 &   -7.41  & 2.74  & -8.30  & 3.48 & -8.37 & 3.50 &
                       -8.45  & 4.29   & -7.65 & 5.56   \\
\multicolumn{2}{c}{QMCI}& 464.5 & -6.65 & 2.75 & -7.73 & 3.48 & -7.59 & 3.53 &
                        -7.85   & 4.31  & -7.13 & 5.57\\
\multicolumn{2}{c}{QMCII}& 685 & -7.83 & 2.74   & -8.50 & 3.48 & -8.58 & 3.52 &
                        -8.61  & 4.30  & -7.74 & 5.56\\
\multicolumn{2}{c}{QHD}  & 500.8 & -7.18  & 2.74  & -8.14  & 3.48  & 
                           -8.19 & 3.50 & -8.30 & 4.29  & -7.57 & 5.56 \\
\multicolumn{2}{c}{Exp.} &     & -7.98  & 2.73 & -8.45 & 3.48 & -8.57 & 3.47 & 
                               -8.66 & 4.27 & -7.86 & 5.50 \\
\hline
\multicolumn{2}{c}{C.M.} &     & -0.67  &  & -0.21 &  & -0.18 &  & 
                               -0.08 &  & -0.03 &  \\
\end{tabular}
\label{tab:two}
\end{table}
%
%%%%%%%%%%%%%%%%%%%%%%%%%%%%%%%%%%%%%%%%%%%%%%%%%%%%%%%%%%%%%%%%%%%
%%%%%%%%%%%%%%%%%%%%%%%%%%%%%%%%%%%%%%%%%%%%%%%%%%%%%%%%%%%%%%%%%%%
%
\begin{table}[tbhp]
\caption{Spin-orbit splittings of the highest occupied proton and neutron
levels in $^{208}{\rm Pb}$.
For the MQMC model the scalar coupling to the quarks is $g_s^q = 3$.}
\medskip
\begin{tabular}[b]{cl|@{\extracolsep{-1pt}}cccccc}
protons & & $ \ \ \ \ \eta = 3$ & $ \eta = 4 $ & QHD 
& QMCI& QMCII & expt.  \cite{CAMPI72} \\
 \cline{3-8}
$\Delta E (2d_{5/2}-2d_{3/2})$ & (MeV) &\ \ \ \  -1.42 & -1.41 & -1.39 
& -0.55 & -1.55 & -1.3 \\
$\Delta E (1g_{9/2}-1g_{7/2})$ & (MeV) &\ \ \ \  -3.40 & -3.40 & -3.43 
& -1.22 & -3.21 &-4.0 \\
\hline
neutrons & & $ \ \ \ \ \eta = 3$ & $\eta = 4 $ & QHD 
& QMCI & QMCII& expt. \cite{CAMPI72} \\
 \cline{3-8}
$\Delta E (3p_{3/2}-3p_{1/2})$ & (MeV) &\ \ \ \  -0.68 & -0.68 & -0.66 
& -0.26 & -0.74 & -0.9  \\
$\Delta E (2f_{7/2}-2f_{5/2})$ & (MeV) &\ \ \ \  -1.80 & -1.78 & -1.74 
& -0.74 & -2.03 & -1.8  \\
\end{tabular}
\label{tab:three}
\end{table}
%
%%%%%%%%%%%%%%%%%%%%%%%%%%%%%%%%%%%%%%%%%%%%%%%%%%%%%%%%%%%%%%%%%%%
%%%%%%%%%%%%%%%%%%%%%%%%%%%%%%%%%%%%%%%%%%%%%%%%%%%%%%%%%%%%%%%%%%%
%
\begin{table}[tbhp]
\caption{Spin-orbit splittings of the highest occupied proton and neutron
levels in $^{16}{\rm O}$.
For the MQMC model the scalar coupling to the quarks is $g_{\rm s}^q = 3$.}
\medskip
\begin{tabular}[b]{cl|@{\extracolsep{-1pt}}cccccc}
protons & & $ \ \ \ \ \eta = 3$ & $ \eta = 4 $ & QHD 
& QMCI& QMCII & expt.  \cite{CAMPI72} \\
 \cline{3-8}
$\Delta E (1p_{3/2}-1p_{1/2})$ & (MeV) &\ \ \ \  -5.16 & -5.18 & -5.27 & -1.87 
& -4.71 & -6.3 \\
\hline
neutrons & & $ \ \ \ \ \eta = 3$ & $\eta = 4 $ & QHD 
& QMCI & QMCII& expt. \cite{CAMPI72} \\
 \cline{3-8}
$\Delta E (1p_{3/2}-1p_{1/2})$ & (MeV) &\ \ \ \  -5.22 & -5.24 & -5.34& -1.88 
& -4.75 & -6.1 \\
\end{tabular}
\label{tab:four}
\end{table}
%
%%%%%%%%%%%%%%%%%%%%%%%%%%%%%%%%%%%%%%%%%%%%%%%%%%%%%%%%%%%%%%%%%%%
%%%%%%%%%%%%%%%%%%%%%%%%%%%%%%%%%%%%%%%%%%%%%%%%%%%%%%%%%%%%%%%%%%%
%
\begin{table}[tbhp]
\caption{Model Summary. For a detailed description see text.}
\medskip
\begin{tabular}[b]{c|ccccccc}
Model & \multicolumn{4}{c}{MQMC} & QMCI & QMCII & QHD\\ 
 \cline{2-5} 
$\eta \backslash g_{\rm s}^q$& $3\backslash 2$ & $3\backslash 3$ & 
$4\backslash 2$ & $4\backslash 3$ & & &\\ 
\hline
Bag constant & $B(\phi)$& $B(\phi)$& $B(\phi)$& $B(\phi)$& 
$B_0$ & $B_0$ & \rule{.25in}{0.01in}\\
$R/R_0$& 1.37 & 1.28 & 1.37 & 1.28 & 0.99 & 0.97 & \rule{.25in}{0.01in}\\
Scal. potential & 
${1\over 2} m^2_{\rm s}\phi^2$&
${1\over 2} m^2_{\rm s}\phi^2$&
${1\over 2} m^2_{\rm s}\phi^2$&
${1\over 2} m^2_{\rm s}\phi^2$&
${1\over 2} m^2_{\rm s}\phi^2$&
\multicolumn{2}{c}{$\Bigl({1\over 2} m^2_{\rm s}\phi^2
+{1\over 3!} \kappa\phi^3+{1\over 4!} \lambda\phi^4\Bigr)$}\\
$M_{\scriptscriptstyle 0}^*/M$ & 0.63 & 0.63 & 0.63 & 0.63 & 0.80 & 
0.63 & 0.63 \\
Binding energies & 1-3$\%$ & 1-3$\%$  & 1-3$\%$ & 1-2$\%$ & 6-9$\%$ & 1-7$\%$ & 1-3$\%$ \\
Spin-orbit splitting & 12$\%$ & 12$\%$ & 12$\%$ & 12$\%$ & 68$\%$ & 18$\%$ &
11$\%$ \\
Charge density & 2.6$\%$ & 2.6$\%$ & 2.5$\%$ & 2.4$\%$ & 3.0$\%$ & 3.7$\%$ &
2.5$\%$ \\
\end{tabular}
\label{tab:five}
\end{table}
\newpage
%
%
%%%%%%%%%%%%%%%%%%%%%%%%%%%%%%%%%%%%%%%%%%%%%%%%%%%%%%%%%%%%%%%%%%%
%%%%%%%%%%%%%%%%%%%%%%%%%%%%%%%%%%%%%%%%%%%%%%%%%%%%%%%%%%%%%%%%%%%
%
%
%%%%%%%%%%%%%%%%%%%%%%%%%%%%%%%%%%%%%%%%%%%%%%%%%%%%%%%%%%%%%%%%%%%
%%%%%%%%%%%%%%%%%%%%%%%%%%%%%%%%%%%%%%%%%%%%%%%%%%%%%%%%%%%%%%%%%%%
%
\section*{Figures}
\global\firstfigfalse
\newpage
\newpage
\newpage
\newpage
\newpage
\newpage
\newpage
\newpage
\newpage
\newpage
%%
%%
%
%%%%%%%%%%%%%%%%%%%%%%%%%%%%%%%%%%%%%%%%%%%%%%%%%%%%%%%%%%%%%%%%%%%
%%%%%%%%%%%%%%%%%%%%%%%%%%%%%%%%%%%%%%%%%%%%%%%%%%%%%%%%%%%%%%%%%%%
%

\begin{thebibliography}{99}
%
\bibitem{GUICHON88} P. A. M. Guichon, Phys.\ Lett.\ {B200} (1988) 235.
%
\bibitem{FLECK90} S. Fleck, W. Bentz, K. Shimizu and K. Yazaki,
                  Nucl. \ Phys. \ {A 510} (1990) 731.
%
\bibitem{SAITO94a} K. Saito and A. W. Thomas, Phys.\ Lett. \ {B327} (1994) 9. 
%
\bibitem{SAITO95b} K. Saito and A. W. Thomas,    
                   Phys. \ Rev. \ C {52} (1995) 2789. 
%
\bibitem{JIN96a} X. Jin and B. K. Jennings, Phys. \ Lett. {B374} (1996) 13.
%
\bibitem{JIN96b} X. Jin and B. K. Jennings, Phys. \ Rev. \ C {54} (1996) 1427.
%
\bibitem{MUELLER97} H. M\"uller and B. K. Jennings, 
                    Nucl. \ Phys. \ {A626} (1997) 966.
%
\bibitem{GUICHON96} P. A. M. Guichon, K. Saito, E. Rodionov and 
                    A. W. Thomas, Nucl. \ Phys. \ {A601} (1996) 349.
%
\bibitem{SAITO96} K. Saito, K. Tsushima and A. W. Thomas, Nucl. \ Phys.\ A
                     {609} (1996) 339.
%
\bibitem{BLUNDEN96} P. G. Blunden and G. A. Miller, Phys. \ Rev. \ C {54}
                    (1996) 359.
%
\bibitem{SAITO97} K. Saito, K. Tsushima and A. W. Thomas, Phys. \ Rev. \ C {55}
                     (1997) 2637.
%
\bibitem{GUICHON97} P. A. M. Guichon, K. Saito and
                    A. W. Thomas, Austral.\ J. \ Phys. {50} (1997) 115.
%
\bibitem{MUELLER98} H. M\"uller, Phys. \ Rev. \ C {57} (1998) 1974.
%
\bibitem{SAITO95a} K. Saito and A. W. Thomas, Phys. \ Rev. \ C {51} 
                   (1995) 2757.
%
\bibitem{TSUSHIMA97a} K. Tsushima, K. Saito and A. W. Thomas, 
                     Phys. \ Lett. \ {B411}, (1997) 9.
%
\bibitem{TSUSHIMA97b} K. Tsushima, K. Saito, J. Haidenbauer and A. W. Thomas, 
                      Nucl.\ Phys.\ {A 630} (1998) 691.
%
\bibitem{JIN97} X. Jin and B. K. Jennings, Phys. \ Rev. \ C {55} (1997) 1567.
%
\bibitem{SEROT97} For an updated status report, see 
                  B. D. Serot and J. D. Walecka, 
                  {\em Int. J. of Mod. Phys. E{\rm 6}} (1997) 515.
%
\bibitem {REINHARD86} P. G. Reinhard, M. Rufa, J. Maruhn, W. Greiner, and J.
                Friedrich, Z. Phys.\ A {323} (1986) 13.
%
\bibitem {FPW87} R. J. Furnstahl, C. E. Price, and G. E. Walker, Phys.\ Rev.\
                C {36} (1987) 2590.
%
\bibitem {FURNSTAHL89} R. J. Furnstahl and C. E. Price, Phys.\ Rev.\ C 
                {40} (1989) 1398.
%
\bibitem {BODMER89} A. R. Bodmer and C. E. Price, Nucl.\ Phys.\ {A 505}
                (1989) 123.
%
\bibitem {GAMBHIR90} Y. K. Gambhir, P. Ring, and A. Thimet, Ann.\ Phys.\ (N.Y.)
                {198} (1990) 132.
%
\bibitem {FURNSTAHL93} R. J. Furnstahl and B. D. Serot, Phys.\ Rev.\ C 
                {47} (1993) 2338.
%
\bibitem {FST97} R. J. Furnstahl, B. D. Serot, and H.-B. Tang, Nucl.\ Phys.\
                  {A 615} (1997) 441.
%
\bibitem {FST96} R. J. Furnstahl, B. D. Serot, and H.-B. Tang, Nucl.\ Phys.\
           {A 598} (1996) 539.
%
\bibitem {BOGUTA77} J. Boguta and A. R. Bodmer, Nucl.\ Phys.\ {A 292}
                (1977) 413.
%
\bibitem{ADAMI93}  C. Adami and G. E. Brown, Phys.\ Rep. \ 234 (1993) 1.
%
\bibitem{SAITO97b} K. Saito, K. Tsushima and A. W. Thomas, 
                   Phys.\ Lett. \ {B406} (1997) 287. 
%                  
\bibitem{HOROWITZ81} C. J. Horowitz and B. D. Serot, Nucl. \ Phys. \
                  {A368} (1981) 503.
%
%\bibitem{RAY79}  L. Ray and P. E. Hodgson, Phys. \ Rev. \ C {20} (1979) 2403.
%
\bibitem{VRIES87}  H. de Vries, C. W. de Jaeger and C. de Vries,
                   Atomic and Nuclear Data tables {36} (1987) 495.
%
\bibitem{FROIS79}  B. Frois, Nuclear physics with electromagnetic interactions,
                   Lecture notes in physics, 
                   eds. H. Arenh\"ovel and D. Drechsel,
                   vol. 108 (Springer, Berlin, 1979), 52.
%
\bibitem {NEGELE70} J. W. Negele, Phys.\ Rev.\ C {1} (1970) 1260.
%
\bibitem{REINHARD89}  P. G. Reinhard, Rep. \ Prog. \ Phys. {52} (1989) 439.
%
\bibitem{CAMPI72}  X. Campi and D. W. Sprung, Nucl. \ Phys. \ {A194} (1972) 401.
%
\end{thebibliography}
\end{document}